%% file: OptimalSweeps-AdamsEtAl-2019.tex
\documentclass{article}
\usepackage{graphicx}
\usepackage{tabls}
\usepackage{afterpage}
\usepackage{cites}
\usepackage{amsmath}
\usepackage{geometry}
\usepackage{amssymb}
\usepackage{epstopdf}
\usepackage{verbatim}
\usepackage{ucs}
\usepackage{datetime}
\usepackage{color}
\usepackage[normalem]{ulem}
\usepackage{nicefrac}

\graphicspath{ {./figures/} , ./ }
\newcommand{\authorHead}      
   {Michael P. Adams, et al.}
\newcommand{\shortTitle}      
   {Optimal Parallel Transport Sweeps}

\begin{document}

%
%

\normalsize

\setlength{\baselineskip}{16.8pt}
\vspace{-3pt}

%
%

\begin{center}
\textbf{\large \\%
PROVABLY OPTIMAL PARALLEL TRANSPORT SWEEPS ON SEMI-STRUCTURED GRIDS \\
}
%
%
\vspace{12pt}
\setlength{\baselineskip}{14pt}
\textbf{Michael P. Adams$^1$, Marvin L. Adams$^1$, W. Daryl Hawkins$^1$, 
\\ Timmie
Smith$^2$, Lawrence Rauchwerger$^2$} \\
$^1$Dept. of Nuclear Engineering;  $^2$Dept. of Computer Science and Engineering \\ 
Texas A\&M University, 3133 TAMU, College Station, TX  77843-3133 \\
\{mpadams, mladams, dhawkins\}@tamu.edu, timmie.smith@gmail.com, rwerger@tamu.edu \\

%
%
\vspace{6pt}
\textbf{Nancy M. Amato}\\
Dept. of Computer Science, University of Illinois  \\
namato@illinois.edu \\
\vspace{6pt}
\textbf{Teresa S. Bailey, Robert D. Falgout, Adam Kunen, Peter Brown}\\
Lawrence Livermore National Laboratory \\
bailey42@llnl.gov; rfalgout@llnl.gov; kunen1@llnl.gov, brown42@llnl.gov\\
%
%

\end{center}

\section*{ABSTRACT}
\begin{quote}
\begin{small}
We have found provably optimal algorithms for full-domain discrete-ordinate
transport sweeps on a class of grids in 2D and 3D Cartesian geometry that
are regular at a coarse level but arbitrary within the coarse blocks.  We describe these algorithms and show that they always execute the full eight-octant (or
four-quadrant if 2D) sweep in the minimum possible number of stages for a given $P_x
\times P_y \times P_z$ partitioning.  Computational results confirm that our
optimal scheduling algorithms execute sweeps in the minimum possible stage
count.  Observed parallel efficiencies agree well with our performance model.
Our PDT transport code
has achieved approximately $68\%$ parallel efficiency with $>1.5M$ parallel threads, relative to 8 threads, on a
simple weak-scaling problem with only three energy groups, 10 direction per octant, and
4096 cells/core.  We demonstrate similar efficiencies on a much more realistic set of
nuclear-reactor test problems, with unstructured meshes that resolve fine geometric details. 
These results demonstrate that discrete-ordinates transport sweeps can be executed with high
efficiency using more than $10^6$ parallel processes.

\emph{Key Words}: transport sweeps, parallel transport, parallel algorithms, PDT, STAPL, performance
models, unstructured mesh

\end{small}
\end{quote}

\setlength{\baselineskip}{14pt}
\normalsize

\input{s1-introduction}

\input{s2-sweeps}

\input{s3-proofs}

\input{s4-optimal}

\input{s5-results}
\input{s6-conclusions}

\section*{ACKNOWLEDGEMENTS}

Part of this work was funded under a collaborative research contract from
Lawrence Livermore National Security, LLC.  Part of this work was performed
under the auspices of the Center for Radiative Shock Hydrodynamics at the
University of Michigan and part under the auspices of the Center for Exascale 
Radiation Transport at Texas A\&M University, both of which have been funded 
by the DOE NNSA ASC Predictive Science
Academic Alliances Program.  Part of this work was funded under a collaborative
research contract from the Center for Exascale Simulation of Advanced Reactors
(CESAR), a DOE ASCR project.  Part of this work was performed under the auspices
of the U.S. Department of Energy by Lawrence Livermore National Laboratory under
Contract DE-AC52-07NA27344.  This research used resources of the Argonne 
Leadership Computing Facility, which is a DOE Office of Science User Facility 
supported under Contract DE-AC02-06CH11357. 

\input{s7-references}

\newpage

\begin{appendix}

\include{s8-appendix1}

\include{s9-appendix2}

\end{appendix}

\end{document}

%% file: s1-introduction.tex
\section{INTRODUCTION} \label{sec:intro}

Deterministic particle-transport methods approximate the particle angular flux
(or density or intensity) in a multidimensional phase space as a function of
time.  The independent variables that define the solution phase space are
position (3 variables), energy (1), and direction (2).  The most widely used
discretizations in energy are \emph{multigroup} methods, in which the solution
is calculated for discrete energy ``groups.''  The most common directional
discretizations are \emph{discrete-ordinates} methods, in which the solution is
calculated only for specific directions.  In the most widely used methods, the
solution for a given spatial cell, energy group, and direction depends only on:
1) the total volumetric source within the cell, and 2) the angular flux for that
group and direction that is incident upon the cell surface.  Each incident flux
is the outgoing flux from an adjacent ``upstream'' cell or is given by boundary
conditions.

To solve the transport equation for the full spatial domain, for a given collection
of energy groups, and for a single direction, one approach is to start with the cell
(or cells) whose incident fluxes for that direction are all provided by boundary
conditions.  (For any direction from a typical quadrature set and a rectangular spatial
domain, this would be one cell at one corner of the domain.)   Once the solution
is found for this cell, its outgoing fluxes complete the dependencies for its
downstream neighbors, whose solutions may then be computed.  Their outgoing
fluxes satisfy their downstream neighbors' dependencies, etc., so each set of
cells that gets completed readies another set, and the computation ``sweeps''
across the entire domain in the direction being solved.  Performing this process
for the full set of cells and directions is called a transport sweep.

The full-domain boundary-to-boundary sweep, in which all angular fluxes in a set
of energy groups are calculated given previous-iterate values for the volumetric
fixed-plus-collisional source, forms the foundation for many iterative methods
that have desirable properties \cite{Adams-Larsen}.  One such property is that
iteration counts do not change with mesh refinement and thus do not grow as
resolution is increased in a given physical problem---an important consideration
for the high-resolution transport problems that require efficient massively
parallel computing.  A transport sweep calculates $\psi^{(l+1/2)}_{m,g}$ via the
numerical solution of:
\begin{equation}
\label{eq:source_iteration}
\vec\Omega_m \cdot \vec \nabla \psi^{(\ell + 1/2)}_{m,g} + \sigma_{t,g}
\psi^{(\ell + 1/2)}_{m,g} = q^{(\ell)}_{tot,m,g} \; \; , \quad \text{all }m, \quad \text{all } g \in \text{ the groupset,}
\end{equation}
where $q^{(\ell)}_{tot,m,g}$ includes the collisional source evaluated using
fluxes from a previous iterate or guess (denoted by superscript $\ell$).  We
emphasize that this is a complete boundary-to-boundary sweep of all directions,
respecting all upstream/downstream dependencies, with no iteration on interface
angular fluxes.

\begin{figure}[!htb]
\centering
\includegraphics[scale=0.25]{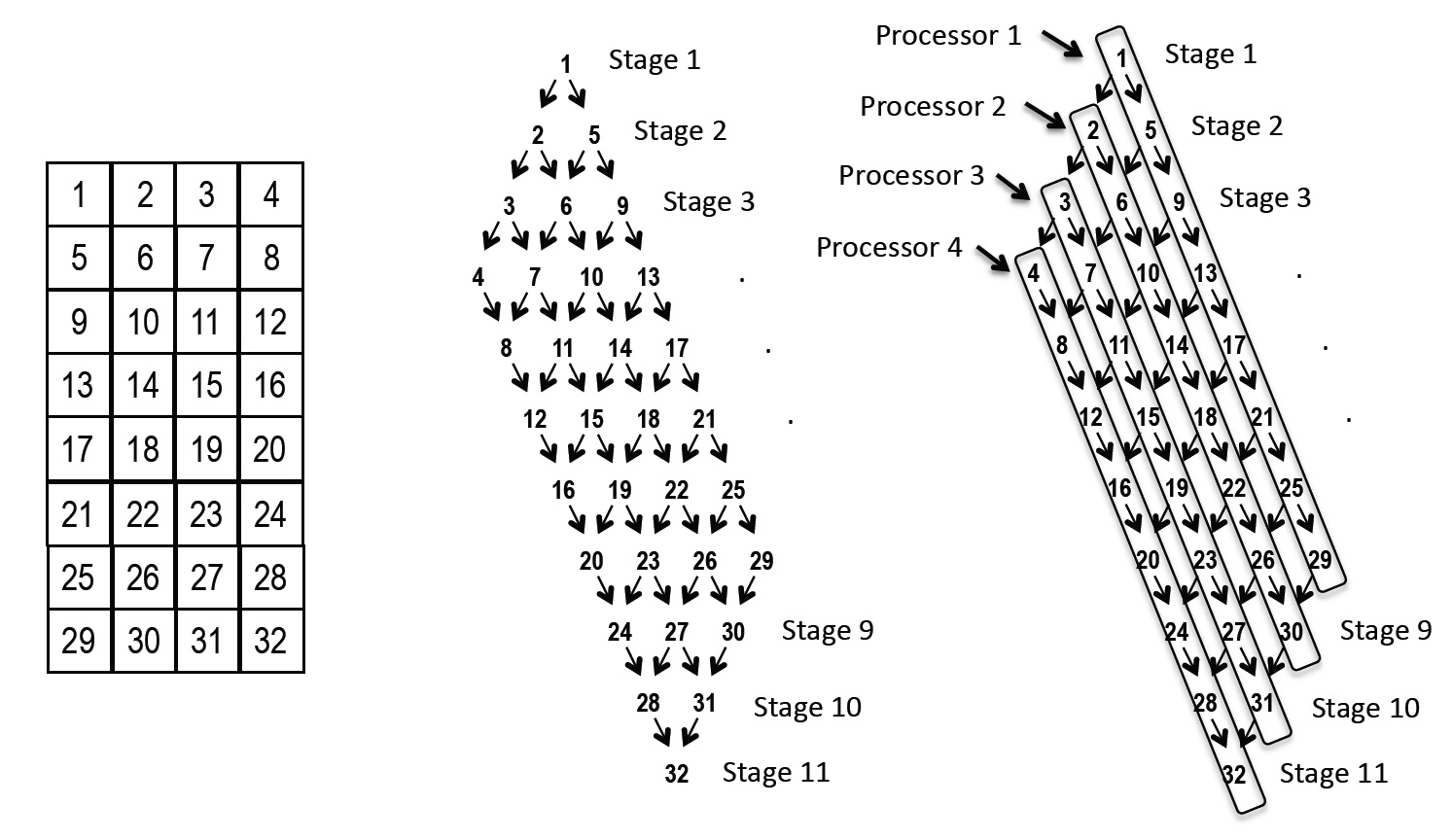}
\caption{(a) Example 2D problem with a spatial grid of $4 \times 8$ cells or
cellsets.  (b) Example TDG for a sweep of a single direction (down and right).
(c) KBA partitioning, with a columns of cellsets assigned to each of four
processors.  Tasks on a given level of the graph can be executed in parallel.}
\label{fig:tdg}
\end{figure}

The parallel execution of a sweep is complicated by the dependencies of cells on
upstream neighbors.  A task dependence graph (TDG) for one direction
in a 2D example (Figs.~\ref{fig:tdg}a and b) illustrates the issue: tasks at a
given level of the graph cannot be executed until some tasks finish on the
previous level.  This originally led to a widespread perception that parallel
sweeps cannot be efficient beyond a few thousand parallel processes and provided
motivation for researchers to seek iterative methods that do not use full-domain
sweeps \cite{denovo,Zerr}.  Such methods offer the possibility of easier
scaling to high process counts---for a single iteration's calculation---but
iteration counts may increase as each process's physical subdomain size
decreases, which tends to happen as resolution and process count both
increase.

In this paper we focus on discrete-ordinates transport sweeps and describe new
parallel sweep algorithms.  We demonstrate via theory, models, and computational
results that our new {\em provably optimal} sweep algorithms enable efficient parallel sweeps
out to O($10^6$) {parallel processes}, even with modest problem sizes 
($O$(1M) cell-energy-direction elements per process).  
We describe a framework for understanding and exploiting the available concurrency in a sweep, recognizing that fundamental dependencies prevent sweeps from being ``embarrassingly parallel.'' We discuss and integrate algorithmic features from past research efforts, providing a comprehensive view of the trade-space available for sweep optimization \cite{KBA,Dorr,ComptonClouse,Pautz,BaileyFalgout}. 

The key components of a sweep algorithm are {\em partitioning} (dividing the domain
among processes), {\em aggregation} (grouping cells, directions, and energy
groups into ``tasks''), and {\em scheduling} (choosing which task to execute if
more than one is available).  
The KBA algorithm devised by Koch, Baker, and Alcouffe \cite{KBA} and the algorithm by Compton and Clouse \cite{ComptonClouse} exploit parallel concurrency enabled by particular partitioning and aggregation choices.  We generalize this as follows.  Given a grid with $N_r$ spatial cells, let us aggregate cells in to $N^{cs}$ brick-shaped cellsets in a $N_x^{cs} \times N_y^{cs} \times N_z^{cs}$ array.  We then distribute these cellsets across processes, with the possibility of assigning more than one cellset to each process.  This corresponds to ``blocks'' in KBA and spatial domain ``overloading'' in other work \cite{ComptonClouse,BaileyFalgout}.  It is possible to also distribute energy groups and/or quadrature directions across different processes, but in this work we focus on spatial decomposition.

In this paper we
limit our analysis to ``semi-structured'' spatial meshes that can be unstructured at a fine level
but are orthogonal at a coarse level, allowing for aggregation into a regular
grid of $N^{cs}_x \times N^{cs}_y \times N^{cs}_z$ brick-shaped cellsets.  Fully
irregular grids introduce complications that we will not address in this paper.
We assume spatial domain decomposition in which each process owns a
contiguous brick-shaped subdomain.  In a future communication we expect to
address decompositions in which a process may own non-contiguous portions of the
spatial domain \cite{mc2015-sweep}.  In the analysis of sweep optimality presented 
below, we assume load-balanced cellsets, with each cellset containing the same 
number of cells with the same number of spatial degrees of freedom. 
The work presented here is based on a recent conference
paper \cite{opt-sweep} but is augmented to include:  1) an extension of our
optimal-sweep algorithm to reflecting boundaries, 2) an improved performance
model, 3) updated and extended numerical results, and 4) a relaxation of
constraints on spatial meshes.

The KBA algorithm devised by Koch, Baker, and Alcouffe \cite{KBA} is the most
widely known parallel sweep algorithm.  KBA {\em partitions} the problem by
assigning a column of cells to each process, indicated by the four diagonal task
groupings in Fig.\ref{fig:tdg}c.  KBA parallelizes over planes logically
perpendicular to the sweep direction---over the breadth of the TDG.  Early and
late in a single-direction sweep, some processes are idle, as in stages 1-3 and
9-11 in Fig.\ref{fig:tdg}.  In this example, parallel efficiency for an {\em
isolated} single-direction sweep could be no better than $8/11 \approx 73\%$.
KBA is much better, because when a process finishes its tasks for the first
direction it begins its tasks for the next direction in the octant-pair that has
the same sweep ordering.  That is, each process  begins a new TDG as soon as it
completes its work on the previous TDG, until all directions in the octant-pair
finish.  This is equivalent to concatenating all of an octant-pair's TDGs into a
single much longer TDG.  This lengthens the ``pipe'' and increases efficiency.
If there were $n$ directions in the octant pair, then the pipe length is
$n\times8$ in this example, and the efficiency would be
$(n\times8)/(3+n\times8)$ if communication times were negligible.

The scheduling algorithm described here is valid for any spatial grid of $N_r$
cells that can be aggregated into $N^{cs} = N^{cs}_x \times N^{cs}_y \times
N^{cs}_z$ brick-shaped cellsets.  A familiar example of a non-orthogonal grid
with this property is a reactor lattice.  As the term ``lattice'' implies, these
grids are regular at a coarse level despite being unstructured at the cell
level.  Additionally, an unstructured mesh that is ``cut'' along full-domain
planes can employ the algorithm described here.  As we describe below, prismatic
grids that are extrusions of 2D meshes into $N_z$ cell-planes---such as those 
commonly found in 3D nuclear-reactor analysis---offer advantages
in optimizing sweeps, but extruded grids are not required by our algorithm.

The coarse regularity of brick-shaped cellsets allows us to {\em partition} the domain into a $P_x
\times P_y \times P_z$ process grid, with $P=$ number of processes $=P_xP_yP_z$.
The work to be performed in the sweep is to calculate the angular intensity for
each of the $N_m$ directions in each of the $N_g$ energy groups in each of the
$N_r$ spatial cells, for a total of $N_mN_gN_r$ fine-grained work units.  The
finest-grained work unit is calculation of a single direction and energy group's
unknowns in a single cell; thus, we describe the sweeps that we analyze here as
use ``cell-based.''  Methods based on solutions along characteristics permit
finer granularity of the computation; in particular, ``face-based'' sweeps are
possible, and with long-characteristic methods ``track-based'' sweeps are
possible.  Face-based and track-based sweeps offer advantages over cell-based
sweeps in terms of potential parallel efficiency, but in this paper we focus on
cell-based sweeps.

We {\em aggregate} fine-grained work units into coarser-grained tasks, with each
task being the solution of the angular fluxes in $A_g$ groups, $A_m$
directions, and $A_r$ spatial cells.  (The $A$s are ``aggregation factors.'')
Since our scheduling algorithm is based on brick cellsets, $A_r$ is constrained
by the level of regularity in the grid.  We use the term ``cell subset'' to
refer to the smallest orthogonal units of the mesh, which we can combine into
cellsets as we see fit.  Thus, if our grid is a lattice of $N^{sub}_x \times
N^{sub}_y \times N^{sub}_z$ brick subsets of $A^{sub}_r$ cells, then $A_r$ will
be an integer multiple of $A^{sub}_r$.  Our choice of ``subset aggregation factors''
$A_x$, $A_y$, and $A_z$ determines our cellset layout, with each
$N^{cs}_u=N^{sub}_u/A_u$.

In order to maintain load balance, we require that each process in our
partitioning scheme own the same number of cellsets $\omega_r \equiv N^{cs}/P$.
Here, $\omega_r$ is the spatial ``overload factor'', and if it is greater than
one we say that our partitioning and aggregation scheme is ``overloaded'', since
processes own multiple cellsets.  This can be broken down as $\omega_r =
\omega_x \times \omega_y \times \omega_z$, with $\omega_u = N^{cs}_u/(P_uA_u)$.
As will be clear from the efficiency formulas in Sec.~\ref{sec:par_sweeps},
there can be significant benefit from overloading.
With everything partitioned and aggregated, each process is responsible for
$\omega_r$ cellsets, $\omega_g \equiv N_g/A_g$ group-sets, and $\omega_m \equiv
N_m/A_m$ direction-sets, for a total of $\omega_m\omega_g\omega_r$ tasks.

The $A_m$ directions that are aggregated together are required to be within the same
octant.  The sweep for directions in a given octant must begin at one of the
eight corners of the spatial domain and proceed to the opposite corner.
If direction-sets from multiple octants are launched at the same time,
there will be ``collisions'' in which a process or set of processes will have
multiple tasks available for execution.  A {\em scheduling} algorithm is
required for choosing which task to execute.

Scheduling algorithms are a primary focus of this paper.
%
%
Our work builds on heuristics-based scheduling algorithms that previous researchers devised \cite{ComptonClouse,Pautz,BaileyFalgout} to address the schedule conflicts that arise from launching simultaneous sweep fronts from all corners of the spatial domain.  In this paper we introduce a family
of scheduling algorithms that execute the complete 8-octant sweep in the minimum
possible number of ``stages,'' where a stage is defined as execution of a single task (cellset/direction-set/groupset) and subsequent communication, by each process that has work available.  We outline a proof of optimality for one
member of the family, discuss the others, and present computational results, which demonstrate that our optimal scheduling algorithms do indeed complete their sweeps in the minimum possible number of stages and provide high efficiency even at high process counts.

With an optimal scheduling algorithm in hand we know how many stages a sweep will
require.  This is a simple function of the partitioning and aggregation parameters chosen for any given problem.  With
stage count known, there is a possibility of predicting execution time via a
performance model, and then using the model to choose partitioning and
aggregation factors that minimize execution time for the given problem on the
given number of processes on the given machine.  The result is what we call an ``optimal sweep
algorithm.''  To recap, the ingredients of the optimal sweep algorithm are:
\begin{enumerate}
    \item A sweep scheduling algorithm that executes in the minimum possible
number of stages for a given problem with given partitioning and aggregation
parameters;
    \item A performance model that estimates execution time for a given problem
as a function of stage count, machine parameters, partitioning, and
aggregation;
    \item An optimization algorithm that chooses the partitioning and
aggregation parameters to minimize the model's estimate of execution time.
\end{enumerate}

In the following section we discuss and quantify key characteristics of parallel sweeps,
including: 1) the idle stages that are inevitable if sweep dependencies are
enforced, and  2) a lower bound on stage count.  We also develop and
discuss simple performance models.  The third section describes our optimal
scheduling algorithms, which achieve the lower-bound stage count found in 
Sec.~\ref{sec:par_sweeps}.  For one algorithm we prove optimality for three kinds of
partitioning: $P_z=1$ (KBA partitioning), $P_z=2$ (``hybrid''), and $P_z>2$
(``volumetric'').  (To simplify the discussion we \emph{define} $x, y, z$ 
such that $P_x \ge P_y \ge P_z$.)  This is the first main contribution of this paper.  In the
fourth section we present our {\em optimal sweep algorithm}, which is made
possible by our optimal scheduling algorithm.  For optimal sweeps, we automate
the selection of partitioning and aggregation parameters that minimize execution
time, as predicted by our performance model, given the knowledge that sweeps
will complete in the minimum possible number of stages for a given set of
parameters.  This is the second main contribution.  Section 5 presents results ranging from 8 to approximately
1.5 million parallel processes, with two different optimal-scheduling algorithms and one
non-optimal algorithm.  In all cases the optimal algorithms complete the sweeps
in the minimum possible number of stages, and performance agrees reasonably well
with the predictions of our performance model.  We offer summary observations,
concluding remarks, and suggestions for future work in the final section.
{Appendices provide graphic illustrations of the behavior of optimally scheduled 
sweeps in 2D and 3D.}

%% file: s2-sweeps.tex
\section{PARALLEL SWEEPS}
\label{sec:par_sweeps}

Consider a $P = P_x \times P_y \times P_z$ process layout on a spatial grid of
$N_r$ cells.  Suppose there are $N_m/8$ directions per octant and $N_g$ energy
groups that can be swept simultaneously.  Then each process must perform
$(N_rN_mN_g)/(P)$ cell-direction-group calculations.  We aggregate these into
tasks, with each task containing $A_r$ cells, $A_m$ directions, and $A_g$
groups.  Then each process must perform $N_\mathrm{tasks}\equiv
\omega_r\omega_m\omega_g = (N_rN_mN_g)/(A_rA_mA_gP)$ tasks.  At each stage at
least one process computes a task and communicates to downstream neighbors.
The complete sweep requires $N_\mathrm{stages}=N_\mathrm{tasks}+N_\mathrm{idle}$
stages, where $N_\mathrm{idle}$ is the number of idle stages for each process.
Parallel sweep efficiency (serial time per unknown / parallel time per unknown
per process) is therefore approximately
\begin{equation} \label{eqn:pareff}
\epsilon = \frac{T_\mathrm{task}N_\mathrm{tasks}}{\left[N_\mathrm{stages}
  \right] \left[T_\mathrm{task}+T_\mathrm{comm}\right] }
=\frac{1}{\left[1+\frac{N_\mathrm{idle}}{N_\mathrm{tasks}} \right]
  \left[1+\frac{T_\mathrm{comm}}{T_\mathrm{task}}\right]} \; ,
\end{equation}
where $T_\mathrm{task}$ is the time to compute one task and $T_\mathrm{comm}$ is
the time to communicate after completing a task. In the second line, the term in
the first [ ] is $1+$ the idle-time penalty and the term in the second [ ] is
$1+$ the comm penalty.  Aggregating into small tasks ($N_\mathrm{tasks}$ large)
minimizes idle-time penalty but increases comm penalty: latency causes
$T_\mathrm{comm}/T_\mathrm{task}$ to increase as tasks become smaller.  This
assumes the most basic comm model, which can be refined to account for
architectural realities (hierarchical networks, random variations, dedicated
comm hardware, latency-hiding techniques, etc.).

In the terms defined above we describe ``basic'' KBA as having $P_z=1$, $A_m=1$ ($\omega_m=N_m$),
$A_g=G$ ($\omega_g=1$), $A_x = N_x/P_x$, $A_y=N_y/P_y$,
and $A_z=$ selectable number of $z$-planes to be aggregated into each task.  (A variant 
described in the original KBA paper is to aggregate directions by octant, which 
means $A_m = N_m/8$ and $\omega_m=8$.)
In
our language, $A_x=N_x/P_x$ and $A_y=N_y/P_y$ translate to
$\omega_x=\omega_y=1$.  With $\omega_z=N_z/(P_zA_z)$, $\omega_m=N_m$ or 8, and
$\omega_g=1$, each process performs $N_\mathrm{tasks} = \omega_m\omega_z$ tasks.
With basic KBA, then, $\omega_z \times \omega_m/4$ tasks (two octants) are pipelined from a
given corner of the 2D process layout in a 3D problem.  For any octant pair the far-corner
process remains idle for the first $P_x+P_y-2$ stages, so a two-octant sweep
completes in $\omega_z \times \omega_m/4+P_x+P_y-2$ stages.  The other three octant-pair
sweeps are similar, so if an octant-pair's sweep does not begin until the
previous pair's finishes, the full sweep requires $\omega_m\omega_z+4(P_x+P_y-2)$
stages.  The parallel efficiency of basic KBA is then
\begin{equation}
\label{eqn:kbae}
\epsilon_{KBA} = \frac{1}{\left[1+\frac{4(P_x+P_y-2)}{\omega_m\omega_z} \right]
  \left[1+\frac{T_\mathrm{comm}}{T_\mathrm{task}}\right]}
\end{equation}
\vspace{-4mm}

KBA inspires our algorithms, but we do not force $P_z=1$ or force particular
aggregation values (such as $A_m=1$ or $A_m=N_m/8$), and \emph{we simultaneously sweep all octants.}
In contrast to KBA, this requires a scheduling algorithm---rules that determine
the order in which to execute tasks when more than one is available.  Scheduling
algorithms profoundly affect parallel performance, as noted in~\cite{BaileyFalgout}.

KBA's choice of $\omega_x = \omega_y = 1$ means that each task completed
satisfies two downstream neighbors' dependencies, which is a substantial
benefit.  As will be seen in Eqs.~(\ref{eqn:nfillxyz}-\ref{eqn:nfill}),
$\omega_x$ and $\omega_y$ values $>1$ cause idle time to 
increase, so it is usually best to set only $\omega_z > 1$.

With basic KBA, the last process to become active is the one that owns the
far corner cellset for a direction.  Since we launch all octants simultaneously,
the last processes to begin computation in our scheme are those at the {\em
center} of the process layout.  The value of this ``pipefill penalty'', the
minimum possible number of stages before a sweepfront can reach the center-most
processes, is
\begin{equation} \label{eqn:nfillxyz}
N_\mathrm{fill} = \omega_x(\frac{P_x+\delta_x}{2}-1)
  + \omega_y(\frac{P_y+\delta_y}{2}-1) + \omega_z(\frac{P_z+\delta_z}{2}-1) ,
\end{equation}
%
where $\delta_u = 0$ or 1 for $P_u$ even or odd, respectively.  If we set
$\omega_x=\omega_y=1$, this becomes
\begin{equation} \label{eqn:nfill}
N_\mathrm{fill} = \frac{P_x+\delta_x}{2}-1 + \frac{P_y+\delta_y}{2}-1 +
  \omega_z(\frac{P_z+\delta_z}{2}-1) .
\end{equation}
Since this is the case in practice, we will use the latter equation in our
efficiency expressions.

Once the central processes begin working, they must complete $N_\mathrm{tasks}$
tasks, which requires a minimum of $N_\mathrm{tasks}$ stages.  Once their last
tasks are completed, there is a pipe emptying penalty with the same value as
$N_\mathrm{fill}$.  As long as dependencies are being respected, then, there is
a hard minimum number of idle stages:
\begin{equation} \label{eqn:nidle}
  N_\mathrm{idle}^\mathrm{min} = 2 N_\mathrm{fill} = P_x+\delta_x-2 +
  P_y+\delta_y-2 + \omega_z(P_z+\delta_z-2) .
\end{equation}
This inevitable idle time then gives us a hard minimum total stage count for a
full-domain sweep:
\begin{eqnarray} \label{eqn:nmin}
N_\mathrm{stages}^\mathrm{min} = N_\mathrm{idle}^\mathrm{min} + N_\mathrm{tasks}
  = P_x+\delta_x-2 + P_y+\delta_y-2 + \omega_z(P_z+\delta_z-2)
    + \omega_r\omega_m\omega_g \;\;
\end{eqnarray}
Important observation:  for a fixed value of $P$,
$N_\mathrm{idle}^\mathrm{min}$ is lower for $P_z=2$ than for the KBA choice of
$P_z=1$, for a given $P$.  In both cases $P_z+\delta_z-2=0$, but with $P_z=2$,
$P_x+P_y$ is lower.  We remark that Eqs.~(\ref{eqn:nfill}-\ref{eqn:nmin}) differ
from those of reference \cite{mc2015-sweep} because here we restrict ourselves to simple $P_x \times
P_y \times P_z$ partitioning, with contiguous spatial subdomains assigned to
each process.

If we could achieve the minimum stage count the optimal efficiency would be:
\begin{eqnarray} \label{eqn:opte}
\epsilon_{opt} = \frac{1}{
  \left[1 + \frac{P_x + \delta_x + P_y - 4 + \delta_y + \omega_z(P_z + \delta_z - 2)}
    {\omega_m\omega_g\omega_z} \right]
  \left[1 + \frac{T_\mathrm{comm}}{T_\mathrm{task}}\right]}.
\end{eqnarray}

It is not obvious that any schedule can achieve the lower bound of Eq.\
(\ref{eqn:nmin}), because ``collisions" of the 8 sweepfronts force processes
to delay some fronts by working on others.  Bailey and Falgout described a
``data-driven" schedule that achieved the minimum stage count in some tests, but
there remained an open question of what conditions would guarantee the minimum
count \cite{BaileyFalgout}.

%% file: s3-proofs.tex
\section{PROOFS OF OPTIMAL SCHEDULING}
\label{sec:proofs}

Here we describe a family of scheduling algorithms that we have found to be
``optimal'' in the sense that they complete the full eight-octant sweep in the
minimum possible number of stages for a given \{$P_u$\} and \{$A_j$\}.  For one such
algorithm---the ``depth-of-graph'' algorithm, which gives priority to the task
that has the longest chain of dependencies awaiting its execution---we sketch
our proof of optimality.  For another---the ``push-to-central'' algorithm, which
prioritizes tasks that advance wavefronts to central planes in the process
layout---we describe scheduling rules but do not prove optimality.  These two
algorithms are endpoints of a one-parameter family of algorithms, each of which
should execute sweeps with the minimum stage count.

To facilitate the discussion and proofs that follow, let us define
\begin{equation}  \label{def_i}
i\in(1,P_x)=\text{the }x\text{ index into the process array},
\end{equation}
with similar definitions for the $y$ and $z$ indices, $j$ and $k$.  We will also
use
\begin{equation}  \label{eq:def_X}
X=\frac{P_x+\delta_x}{2} \;, \;\;\; Y=\frac{P_y+\delta_y}{2} \;, \;\;\; Z
= \frac{P_z + \delta_z}{2} \;
\end{equation}
to define ``sectors'' of the process array, e.g. ($i\in(1,X), \; j\in(1,Y)$) is a sector.
We will use superscripts to represent octants/quadrants, e.g.  $^{++}$ to denote
($\Omega_x>0, \; \Omega_y>0$), $^{-+-}$ to denote ($\Omega_x<0, \; \Omega_y>0,
\; \Omega_z<0$), etc.

The depth-of-graph algorithm is essentially the same as the ``data-driven''
schedule of Bailey and Falgout \cite{BaileyFalgout}, with the exception
of tie-breaking rules, which we find to be important.  The behavior of the algorithm
will become clear in the proofs
that follow.  

The push-to-central algorithm prioritizes tasks according to the
following rules.
\begin{enumerate}
  \item{If $i \le X$, then tasks with $\Omega_x > 0$ have priority over tasks
    with $\Omega_x < 0$, while for $i>X$ tasks with $\Omega_x < 0$ have priority.}
  \item{If multiple ready tasks have the same sign on $\Omega_x$, then for $j\le
    Y$ tasks with with $\Omega_y > 0$ have priority, while for $j>Y$ tasks with
    $\Omega_y<0$ have priority.}
  \item{If multiple ready tasks have the same sign on $\Omega_x$ and $\Omega_y$,
    then for $k\le Z$ tasks with $\Omega_z>0$ have priority, while for $k>Z$ tasks
    with $\Omega_z<0$ have priority.}
\end{enumerate}
Note that this schedule pushes tasks toward the $i=X$ central process plane
with top priority, followed by pushing toward the $j=Y$ (second priority) and
$k=Z$ (third priority) central planes.

The depth-of-graph and push-to-central algorithms differ only in regions of the
process-layout domain in which the ``depth'' priority differs from the
``central'' priority for some octants.  In those regions for those octants, one
can view the two algorithms as differing only in the degree to which they allow
the two opposing octants' tasks to interleave with each other.  The
push-to-central algorithm maximizes this interleaving while the depth-of-graph
algorithm minimizes it.  One can vary the degree of interleaving between these
extremes to create other scheduling algorithms.  Our analysis (not
shown here) indicates that each of these algorithms achieves the minimum possible
stage count.

%
\subsection{Depth-of-Graph Algorithm:  General}
%

The essence of the depth-of-graph scheduling algorithm is that each process
gives priority to tasks with the most downstream dependencies, or the greatest
remaining \emph{depth of graph}.  (By ``graph'' we mean the task dependency
graph, as pictured in Fig.~\ref{fig:tdg}.)  This quantity, which we will denote
$D(O)$ for an octant $O$, is a simple function of 
cellset location and octant direction.  The depth-of-graph algorithm
prioritizes tasks according to the following rules.
\begin{enumerate}
  \item{Tasks with higher $D$ have higher priority.}
  \item{If multiple ready tasks have the same $D$, then tasks with $\Omega_x>0$
    have priority.}
  \item{If multiple ready tasks have the same $D$ and the same sign on
    $\Omega_x$, then tasks with $\Omega_y>0$ have priority.}
  \item{If multiple ready tasks have the same $D$ and the same sign on
    $\Omega_x$ and $\Omega_y$, then tasks with $\Omega_z>0$ have priority.}
\end{enumerate}
We will develop our proof with the aid of indexing algebra, but the core concept
stems from Eq.~(\ref{eqn:opte}).  The formula for $\epsilon_{opt}$ implies that
three conditions are sufficient for a schedule to be optimal:
\begin{enumerate}
  \item The central processes must begin working at the earliest possible
    stage.
  \item The highest priority task must be available to the central processes
    at every stage (i.e., once a central process begins working, it is not
    idle until all of its tasks are completed).
  \item The final tasks completed by the central processes must propagate
    freely to the edge of the problem domain.
\end{enumerate}
If these three criteria are met, a schedule will be optimal as defined by
Eq.~(\ref{eqn:opte}).  For $P_z=1$ and $P_y>1$ the four central processes are
defined by $i \in (X,X+1)$ and $j \in (Y,Y+1)$.  For $P_z>1$ the eight central
processes are defined by these $i$ and $j$ ranges along with $k \in (Z,Z+1)$.

The ``corner'' processes begin at the first stage.  This leads to satisfaction
of the first condition, for the four or eight sweep fronts (for $P_z=1$ or $>1$)
proceed unimpeded to the four or eight central processes, with no scheduling
decisions required.  The second condition is not obvious, but we will
demonstrate that the depth-of-graph prioritization causes it to be met.  Any
algorithm that satisfies the second item will likely achieve the third.  We show
that depth-of-graph does.

We will examine the behavior of the depth-of-graph scheduling algorithm within
three separate partitioning schemes.  The first, $P_z=1$, uses the same
partitioning as KBA; however, as mentioned, we do not impose the same
restrictions on our aggregation, and we launch tasks for all octant-pairs
simultaneously.  The second uses $P_z=2$, which we call the ``hybrid''
decomposition since it shares traits with both the $P_z=1$ case and the $P_z>2$
case.  We call the latter ``volumetric'', since it decomposes the domain into
regular, contiguous volumes.


Since the basic scheme of our algorithm sets priorities based on
\emph{downstream} depth of graph, we will use $D(O)$ to represent this quantity
for octant (or octant-pair) $O$:
\begin{equation}  \begin{array}{cccc}
  D(+-)  =  & (P_x-i)  & + (j-1)   & \\
  D(-+-) =  & (i-1)    & + (P_y-j) & + (k-1)  \\
  \end{array}  \; .
\end{equation}

Since much of the algebra for stage counts depends on the depth of a task
\emph{into} the task graph, we define a direction-dependent variable $s$:
\begin{equation}
  s^{--+} = (P_x-i) + (P_y-j) + (k-1)
\end{equation}
etc.  These are measures of upstream ($s$) and downstream ($D$) dependence chains and
are related by $D(O) + s(O) = $ total depth of graph $ -1= P_x + P_y + P_z - 3$.
We find that $D$ is convenient for discussing priorities, and $s$ is
convenient for quantifying the stage at which a task will be executed.

Our aggregation factors determine what we cluster together as a single task.  A
task is the computation for a single set of $A_r$ cells, for a single set of
$A_g$ energy groups, for a single set of $A_m$ angles.  We define $M\equiv$
the number of tasks per process per quadrant for $P_z=1$ and per octant for
$P_z>1$.  This is different from $N_\mathrm{task}$ discussed above; it takes 1/4
the value if $P_z=1$ and 1/8 otherwise.  We use $m^O\in(1,M)$ to represent a
specific task from the ordered list for octant (or quadrant) $O$, and $\mu$ to
represent the stage at which a task is completed.  Thus, $\mu(m^O,i,j)$ is the
stage at which process $(i,j)$ performs task $m$ in octant $O$.

%
\subsubsection{Sector symmetry}
%

If $P_x$, $P_y$ and $P_z$ are all even, then the sectors are perfectly symmetric
about the planes $i = X + \textonehalf$, $j = Y + \textonehalf$, and $k = Z +
\textonehalf$ (half integer indices denote process subdomain boundaries).  In
the case of $P_u$ odd, there is an asymmetry:  the sector of greater $u$ is one
process narrower.

Equation (\ref{eqn:opte}) shows that the optimum number of stages for an odd
$P_x$ or $P_y$ (or $P_z>2$) equals that for $P_u + 1$.  To simplify the analysis
we will convert cases with any odd $P_u$ to even cases with $P_u + 1$ (except
for $P_z=1$) by imagining additional ``ghost processes." The ``ghost
processes'' do not change the optimal stage count, and they leave us with
perfectly symmetric sectors.  Thus, we assume that $\delta_x = \delta_y = 0$
(and $\delta_z=0$ for $P_z>2$), we focus on the sector with $i \in (1,X)$, $j
\in (1,Y)$ and $k \in (1,Z)$, and we know that other sectors behaves
analogously.


\subsection{$P_z=1$ Decomposition}
\label{sec:2D}


For this partitioning we aggregate such that $\omega_x=\omega_y=1$.  In the next
section, we will discuss how $\omega_z$ and $\omega_m$ are optimized based on a
performance model, but for now they are treated as free variables.  We have
defined $M = \omega_z\omega_g (\omega_m/4)$, which encapsulates the multiple
cellsets, group-sets, and direction-sets within an octant for a process.
Since the values $\omega_x=\omega_y=1$ ensure that every completed task will
satisfy dependencies downstream, our analysis will not directly involve
aggregation factors.

The foundation of the scheduling algorithm we are analyzing is downstream depth
of graph.  $D(O)$ depends on angleset direction and 
cellset location, and different regions of the problem domain will have
different priority orderings.  These regions can be determined by index
algebra.

\subsubsection{Priority regions}

Since we assign priorities to octant-pairs (quadrants) based on $D(O)$, which is a simple function
of $i$ and $j$, it is a simple matter to determine in advance a process's
priorities.  The domain thus divides into
distinct, contiguous regions with definite priorities.  For example, process
$(1,1)$ executes tasks by quadrant in the order $++$, $+-$, $-+$, $--$.  At times we will
find it convenient to refer to a region by its priority ordering.  It is also
convenient to refer to a quadrant as a region's primary, secondary, etc., priority.

The boundaries between regions of different priorities are planes (or lines in
our 2D process layout) defined by the solutions of the equations $D(O_1) =
D(O_2)$ for distinct octants (or quadrants) $O_1$ and $O_2$.  For example, let
us examine quadrants $++$ and $+-$.
\begin{align}  \label{eq:j_eq_Y}
D(++)  & =  D(+-)  \notag \\
\implies (P_x - i) + (P_y - j)  & =  (P_x - i) + (j-1)  \notag \\
\implies j  & =  Y + \frac{1}{2}
\end{align}

(We continue to assume that $P_x$ and $P_y$ are even.) The non-integer value
means the plane passes \emph{between} processes.  Thus, the even case cleanly
divides the problem domain into two regions:  $j<Y$, where $++$ quadrants have
priority, and $j>Y$, where $+-$ quadrants have priority.  Thus, there are no
ties to break for any process for these two quadrants.

Note that Eq.~(\ref{eq:j_eq_Y}) is also the solution of $D(-+) = D(--)$.  There
is an analagous plane bounding the two quadrant pairs with differing signs in
$\Omega_x$, given by
\begin{equation}
i = X + \frac{1}{2}  .
\end{equation}
There are also two quadrant pairs with sign differences in both $\Omega_x$ and
$\Omega_y$.  Solving for these boundaries, we find
\begin{equation}  \label{eq:i_plus_j}
(i-X) + (j-Y) = 1\;\;\; ;  \;\;\;
(i-X) - (j-Y) = 0  .
\end{equation}
We see that the first Eq.~(\ref{eq:i_plus_j}) is a line of slope $-1$ through
the center of the domain, and integer values of $i$ and $j$ satisfy the
equation.  Thus, with $P_u$ both even, there are ``diagonal lines" of processes
for which pairs of octants have the same priority---a tie.  We use a simple
tie-breaking scheme here:  the first tie-breaker goes to tasks with
$\Omega_x>0$, and the second tie-breaker is for $\Omega_y>0$.  Once we apply our
tie-breaker, the diagonal-line processes can be thought of as belonging to the
region that prioritizes the winning quadrant.

\begin{figure}[!htb]
\centering
\includegraphics[scale=0.3]{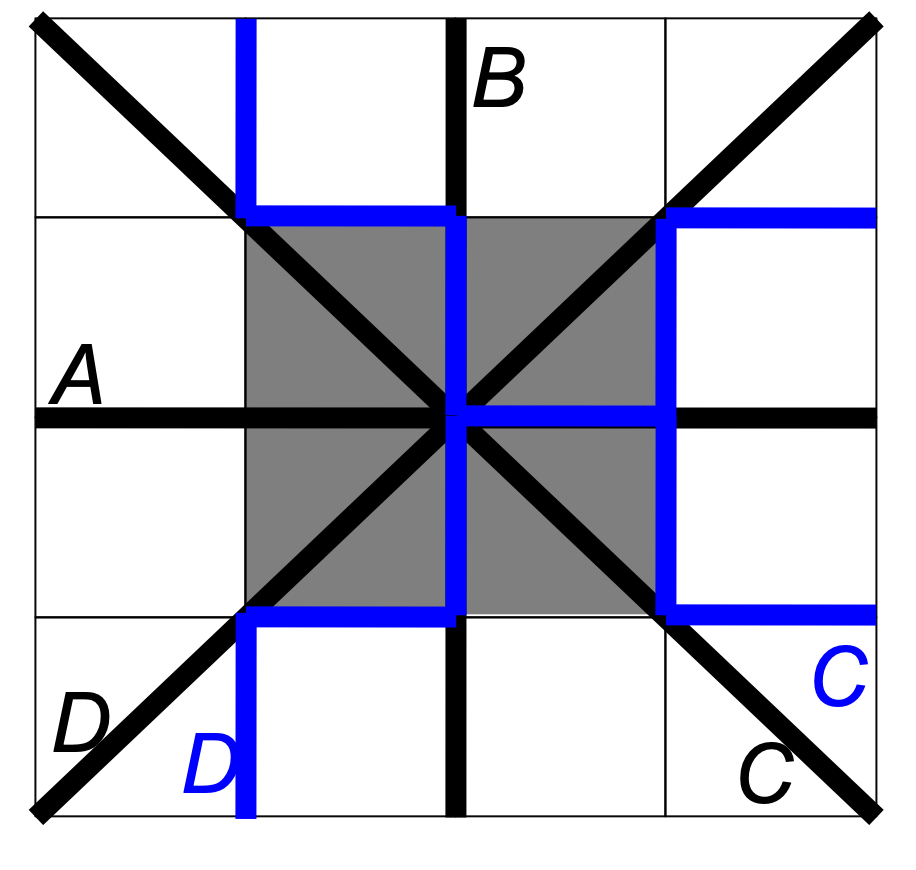}
\caption{{\bf Priority regions} for $P_x$ and $P_y$ even.  Lines $A$ through $D$
represent Eqs.\ (\ref{eq:j_eq_Y}-\ref{eq:i_plus_j}), respectively, with blue
versions after tie-breaker.  Central processes are shaded.  The figure is
``zoomed in'' on the central portion of the process domain, which may have
arbitrarily large extent.}
\label{fig:2d_regions}
\end{figure}

Figure~\ref{fig:2d_regions} shows the central portion of a problem domain
divided into priority regions.  We call attention to ``central'' processes,
shaded in the figure, which determine much of the behavior of our scheduling
algorithm.  Note that the process layout in the figure could be either the
entire domain or a small central subset; the lines and all they signify are the
same either way.

\subsubsection{Primary quadrant:  filling the pipe}
\label{sec:p1_quad}

At the outset of a sweep only the four ``corner" processes have their incoming
fluxes (from boundary conditions).  Each corner process completes its first
task at stage one, which satisfies dependencies for its downstream neighbors.
Thus begin the waves of task-flow called the sweep.

Let us examine the order in which processes complete tasks in their primary
quadrant (e.g., quadrant $++$ for sector $--$).  We begin at stage $\mu=1$ with
process $(i,j)=(1,1)$ performing task $m=1$.  Once this is completed (i.e., in
stage 2), processes $(1,2)$ and $(2,1)$ can perform task $1$, and process
$(1,1)$ moves on to task 2.  In stage 3, processes $(1,3)$, $(2,2)$, and
$(3,1)$ perform task 1, processes $(1,2)$ and $(2,1)$ perform task 2, and
process $(1,1)$ performs task 3.  We can generalize this pattern with the
simple expression
\begin{equation}
  \label{sector_one_primary}
  \mu(m^{++}, i, j) = (i-1) + (j-1) + m^{++} = s^{++} + m^{++}.
\end{equation}

For a given process ($i,j$ pair), this equation describes the task number
incrementing with each successive stage.  For a given task (value of $m^{++}$),
it describes a set of processes along a line of slope $-1$ moving up and right
at each stage.

The procession of tasks proceeds in this way from each corner, with the
processes in each sector performing tasks from their primary quadrant as long
as they last.  Thus, we find that primary quadrant task execution follows
Eq.~(\ref{sector_one_primary}) as well as the analogous:
\begin{equation}
  \label{sector_two_primary}
  \mu(m^{+-}, i, j) = (i-1) + (P_y-j) + m^{+-} = s^{+-} + m^{+-},
\end{equation}
\begin{equation}
  \label{sector_three_primary}
  \mu(m^{-+}, i, j) = (P_x-i) + (j-1) + m^{-+} = s^{-+} + m^{-+},
\end{equation}
and
\begin{equation}
  \label{sector_four_primary}
  \mu(m^{--}, i, j) = (P_x-i) + (P_y-j) + m^{--} = s^{--} + m^{--}.
\end{equation}


\subsubsection{Starting on the central processes}

As can be seen from Eqs.~(\ref{eq:j_eq_Y})-(\ref{eq:i_plus_j}) or
Fig.~\ref{fig:2d_regions}, each quadrant has top priority for one entire sector,
so even if other tasks are available, these stage counts will hold within the
initial sector.  Thus, the central processes are reached in $X+Y-1$ stages,
just as in Eq.~(\ref{eqn:nfill}), which satisfies the first condition for
optimality:  the central processes begin work at the first possible stage.

This result also gives us a start on the second condition:  the central
processes stay busy until their work is done.  It is clear from
Eqs.~(\ref{sector_one_primary}-\ref{sector_four_primary}) that successive tasks
in a given quadrant take place at successive stage counts.  If the central
process gets its first task at stage $\mu$, it will receive the second at
$\mu+1$, etc.  This guarantees that all of the tasks in a central process's
highest priority quadrant will arrive in sequence, allowing the process to
stay busy as it processes its first quadrant.

Observe the symmetry between sectors:  as process $(X,Y)$ computes its $++$
tasks, $(X,Y+1)$ and $(X+1,Y)$ compute their $+-$ and $-+$ tasks, respectively,
and communicate to $(X,Y)$.  This satisfies half of $(X,Y)$'s dependencies for
these two quadrants.  The other dependencies remain; for example, tasks in the
$+-$ quadrant cannot be executed at $(X,Y)$ until $(X-1,Y)$ has computed them
too.  This is addressed next.


\subsubsection{Second-priority tasks}

The second-priority quadrant for process $(X,Y)$ is $+-$.  Its tasks began at
$(1,P_y)$ and propagated as a mirror image of the $++$ tasks from $(1,1)$.  The
first $+-$ task becomes available to process $(1,Y)$ at stage
\begin{equation}
\mu = (1 - 1) + (P_y - Y ) + 1 = Y + 1 \; ,
\end{equation}
just as the first $++$ task becomes available to $(1,Y+1)$.  However, $(1,Y)$
works on $++$ tasks until they are exhausted; only then will it begin the $+-$
tasks, whose dependencies are already satisfied (one
by boundary conditions, one by information from $(1,Y+1)$).  This results in a delay on
secondary-quadrant tasks (e.g., quadrant $+-$ for sector $--$ processes), given by
\begin{equation*}
d = \text{delay} = \mu(1+M^{++}, 1, Y) - \mu(1^{+-}, 1, Y)  = M - 1 \; .
\end{equation*}

As the tail of the $++$ task wave propagates along the processes at $j = Y$,
the first $+-$ task flows in right behind it.  This is illustrated in the fifth frame of Fig.~\ref{fig:2d-sweep} in App.~\ref{sec:appendix1}.  The wave-front propagates
as as a mirror image of the primary quadrant, starting from $(1,Y)$.  The
dependencies from $j = Y+1$ have already been satisfied, those at the boundary
are given, and the final $++$ task has already swept past, so we find that
\begin{equation}
\mu = (i-1) + (P_y-j) + (M-1) + m^{+-}
\end{equation}
for processes that give quadrant $+-$ second priority.  This includes the central
process, which thus transitions smoothly from $++$ task to $+-$ tasks, staying
busy until tasks from these two quadrants are finished.


\subsubsection{Third-priority tasks}

Beginning at $(X,1)$, the central process's third-priority quadrant
($-+$) begins its march through the ($--$) sector with a progression symmetric 
to that of the central process's second-priority quadrant (discussed in the immediately 
preceding subsection), with
\begin{equation}
\mu = (P_x-i) + (j-1) + (M-1) + m^{-+} \;,
\end{equation}
for the processes that give second priority to $-+$ second---the processes that 
own cellsets below the diagonal in the $(--)$ sector.  This region stops just
shy of the central process; its boundary is given by Eq.~(\ref{eq:i_plus_j}).

The second- and third-priority task waves arrive at the processes given by the
equal-depth equation at the same stage---the processes that own cellsets on 
the diagonal---but the third-priority tasks lose the
tie-breaker.  On each side of the boundary, processes continue to execute
their second-priority tasks as the third-priority tasks become available.

The central process (and the others along the diagonal line) finishes the last
second-priority quadrant's task at stage
\begin{align*}
\mu & = (X-1) + (P_y-Y) + (M-1) + M^{+-}  \\
    & = X + Y + 2M - 2 \; .
\end{align*}
The processes across the boundary finished their second-priority tasks the
stage before.  Now that the second-priority tasks are finished, the processes
on both sides begin their third-priority tasks.  The central process, $(X,Y)$,
has had the dependency from $(X,Y-1)$ met from the time it began its $+-$ tasks;
it simply prioritized the latter tasks over the available $-+$ work.  Recalling our sector
symmetry, the dependency from $(X+1,Y)$ was met as it completed its first $++$
task.  Thus, the central processes all stay busy through their third-priority
quadrants.

The two incoming task waves (from the second- and third-priority quadrants) arrived at the (tie-broken jagged-diagonal) region boundary intact.  Each process adjacent to the region boundary executes all of its second-priority tasks.  When the last second-priority task is complete, the two regions begin their third-priority tasks all along the (jagged-diagonal) region boundary.
Thus, the waves resume their propagation delayed but unbroken.  For
$-+$ tasks,
\begin{equation}  \label{eq:mu_second}
\mu = (P_x-i) + (j-1) + (2M-1) + m^{-+} \; .
\end{equation}
Since the $-+$ tasks lose the tie-breaker and are held up a stage earlier, the
$+-$ tasks actually continue their progress a stage earlier:
\begin{equation}  \label{eq:mu_third}
\mu = (i-1) + (P_y-j) + (2M-2) + m^{+-} \; .
\end{equation}
Both task waves sweep along unimpeded, as they now have the highest priority in
their current regions.  As they go, they are fulfilling (in advance)
dependencies for the adjacent sectors' fourth-priority tasks.


\subsubsection{Fourth-priority tasks}

We have seen that each central process begins its first task at the
earliest possible stage.  We have also seen that its supply of tasks is
continuous through its third-priority quadrant.  The dependencies for its
fourth-priority quadrant are the two neighboring central processes.  Since each
fourth-priority quadrant was a higher priority for the other processes, and
since they have all completed their first three quadrants, they have now
satisfied each others' dependencies.  Thus, the second condition for an optimal
schedule has been fulfilled.

The third condition is that the final tasks propagate without delay to the
problem boundaries.  Since the third-priority task waves are already retreating
from the central processes, as shown by Eqs.\
(\ref{eq:mu_second}-\ref{eq:mu_third}), we know that there are no competing
tasks remaining.  These equations also demonstrate that the fourth-priority
dependencies have already been satisfied.  Just as we have seen task waves begin
propagating from $(1,1)$, $(1,Y)$ and $(X,1)$, the fourth-priority wave now
propagates smoothly from $(X,Y)$, with
\begin{equation}
\mu = (P_x-i) + (P_y-j) + (3M-1) + m^{--}  \; .
\end{equation}

The final task of the fourth-priority quadrant will thus be executed by
process $(1,1)$ at stage
\begin{align}
\mu & = (P_x-1) + (P_y-1) + (3M-2) + M^{--}  \\
    & = P_x + P_y + 4M - 4 = P_x + P_y - 4 + N_\mathrm{tasks}  \; ,
\end{align}
which is exactly the minimum we established in Eq.~(\ref{eqn:nmin}).


\subsection{$P_z=2$ Decomposition (``Hybrid'')}
\label{sec:hybrid}

Now consider the case of $P_z=2$.  Above, we considered task groups in terms of
quadrants, which are actually sets of two octants.  We did not specify the
ordering of tasks within a quadrant because the proof holds true regardless of
that order.  Thus, we are free to do the entire ``upward'' ($\Omega_z>0$) octant
first, followed by the entire downward octant, and all of the properties we have
established above are unchanged.

The depth-of-graph algorithm schedules tasks for the $k=1$ processes exactly
this way, and the $k=2$ processes mirror the ordering.  While the lower
processes solve their upward tasks, the upper processes solve their downward
tasks, so that by the time one is done, the other is waiting.  All other
scheduling concerns are handled exactly the same as in $P_z=1$.

This leads to a result that may not be obvious:  if we take a $P_z=1$ problem and
double both $N_z$ and $P_z$, then the task flow for the $k=1$ processes
is indistinguishable from the $P_z=1$ case with the original $N_z$.  They work their $\Omega_z>0$ octants, as
before, and then their $\Omega_z<0$ octants, just as before.  Thus,
using the hybrid decomposition instead of the $P_z=1$ allows for doubling the number
of cells in $z$ and doubling the number of processes \emph{with no increase in
solve time}.

This benefit rests on the initial step of computation.  In $P_z=1$, only four
processes had tasks with no unsatisfied dependencies; now all eight corner
processes launch their primary octants at once.  We experience the same
pipe-fill penalty as before, and the number of tasks per process is the same.
We  perform twice the work with twice the processors in the same time, 
except for a possible communication delay between upper and lower 
halves of the problem.


\subsection{$P_z>2$ Decomposition with $\omega_x=\omega_y=\omega_z=1$ (``Volumetric'')}

In this section we examine what we call a volumetric decomposition, which
is the full extension of the decomposition into three dimensions.  The
same requirements for optimality apply, and the task flow follows the same
principles.  For this analysis, we will assume that $\omega_x=\omega_y=\omega_z=1$, so that
each process owns only a single cellset.

\subsubsection{Priority regions}

Much as before, the domain is divided into regions with different priority
orders based on relative depths of graph for different octants.  For $P_z=1$,
there were six distinct pairs of colliding quadrants (as in
Eqs.~\ref{eq:j_eq_Y}-\ref{eq:i_plus_j}) and eight regions of different priority
orderings (as shown in Fig.~\ref{fig:2d_regions}).  For $P_z>2$, there are 28 
distinct pairs of colliding octants ($7+6+...+1 = 28$), and, as
we will see, 96 different regions with different priority orderings.  
The regions are separated by the planes along
which two octants have equal depth of graph.

\begin{figure}[!htb]
\centering
\includegraphics[scale=0.2]{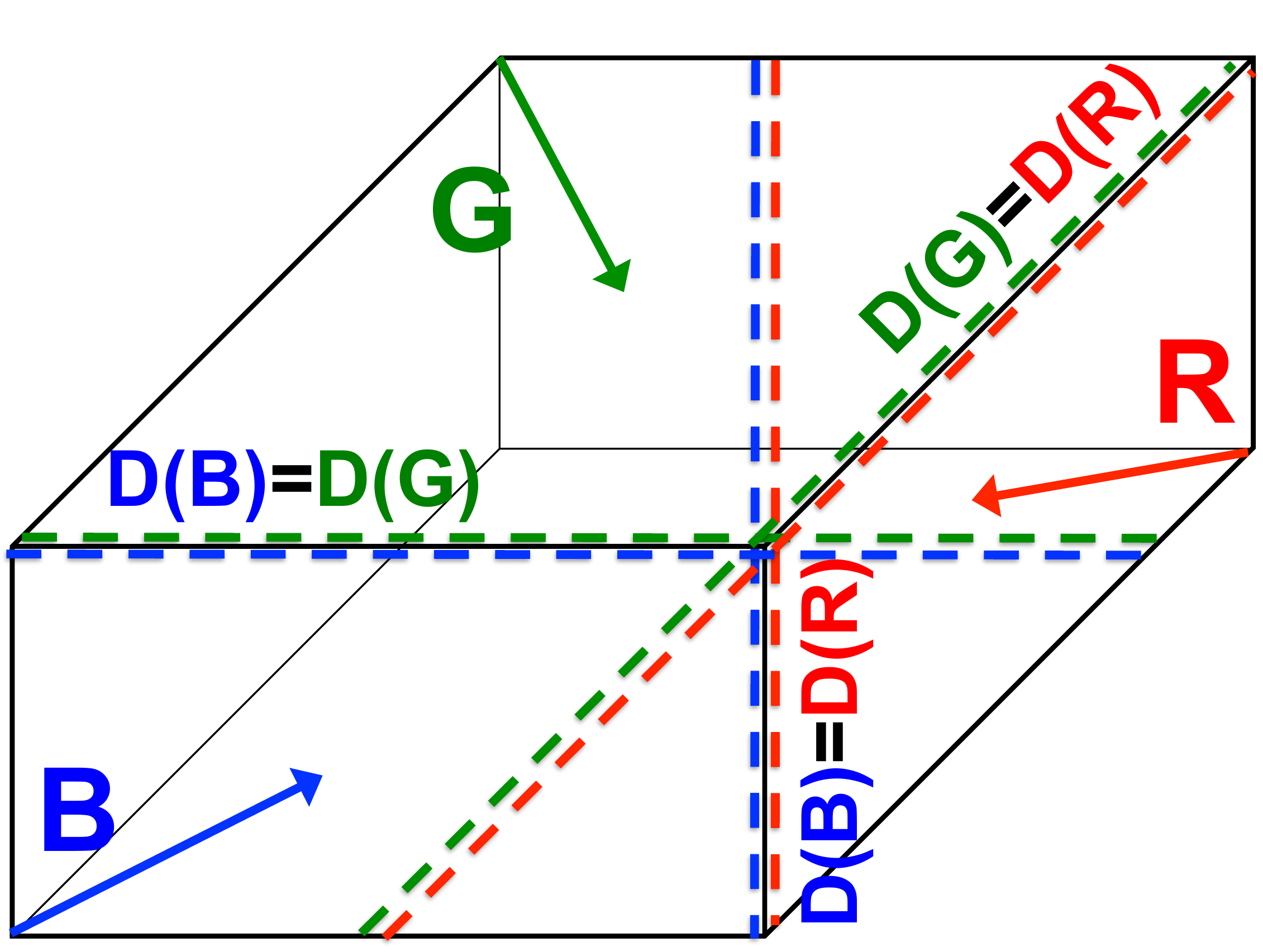}
\includegraphics[scale=0.2]{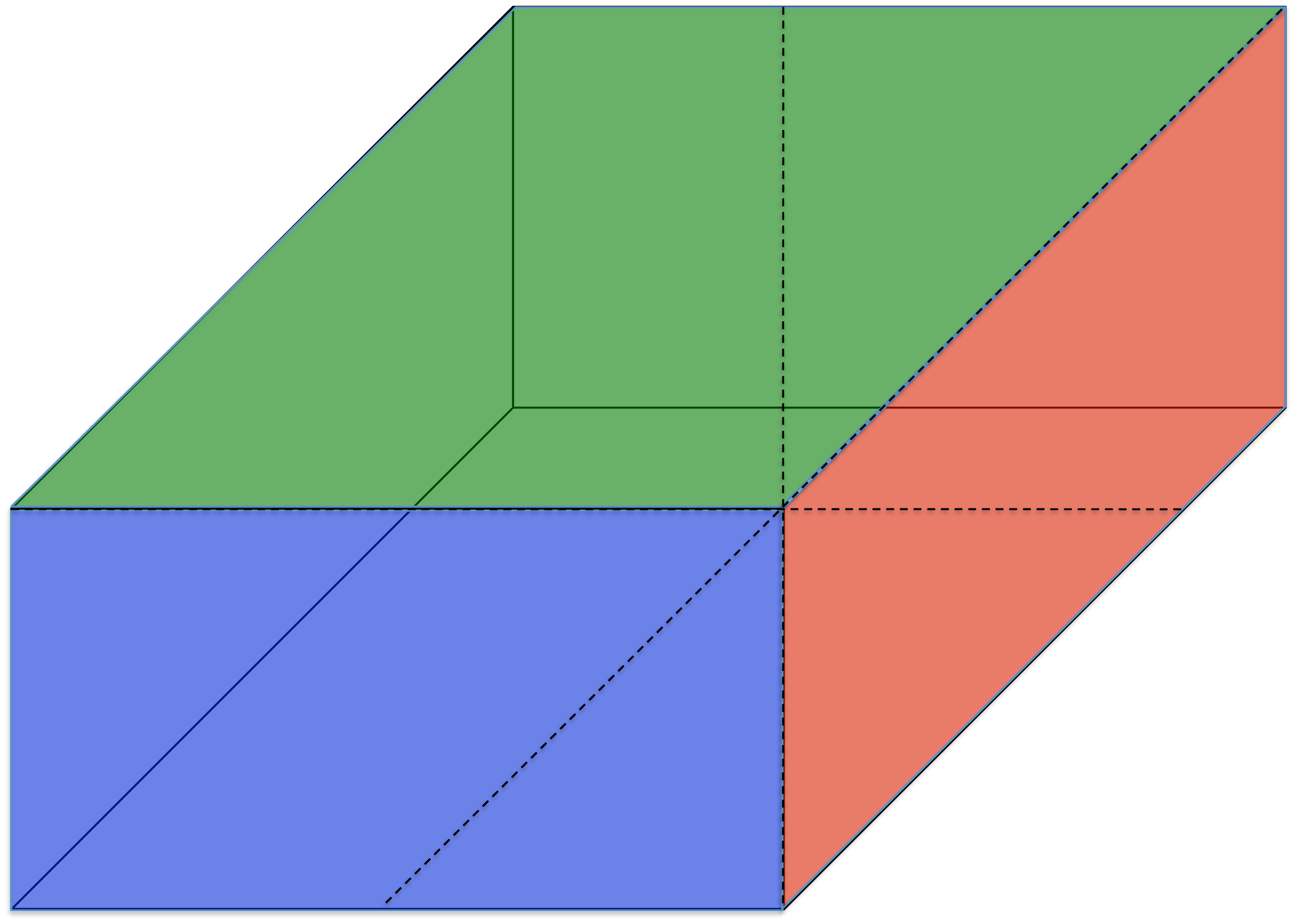}
\includegraphics[scale=0.2]{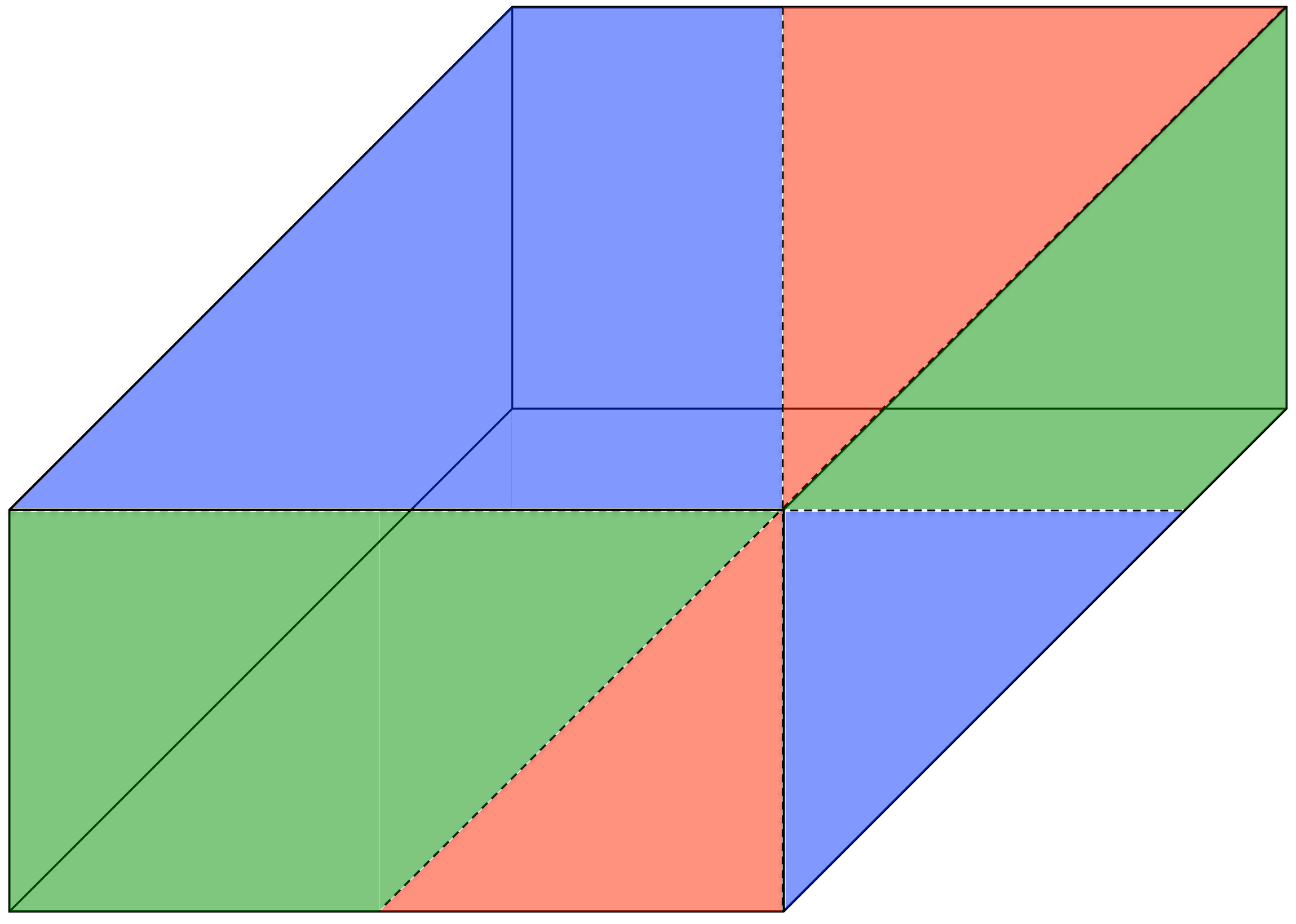}
\caption{Left: Planes of equal depth of graph divide a sector into regions of
different priority.  Center: Each of the \{R,G,B\} octants has second priority in two adjacent regions 
of the six regions.  Right: Each of the \{R,G,B\} octants has third priority in two non-adjacent regions.}
\label{fig:rgb}
\end{figure}

For $P_z=1$ or $P_z=2$, after the primary quadrant there were two quadrants
advancing in each sector.  For $P_z>2$ there are three octants entering each
sector, which we will nickname $R$, $B$ and $G$ (for red, blue and green).  We
will call the primary octant $P$.  The priority regions $(P,R,...)$, $(P,B,...)$
and $(P,G,...)$ are defined by planes that we will call the $RB$, $RG$ and $BG$
boundaries, given by $D(R)=D(B)$, etc., as illustrated in Fig.~\ref{fig:rgb}.
Because these three planes intersect along a single line (perpendicular to the
plane of the figure), they divide the sector into six regions.  (Later we will see that 
other octants further divide the six into twelve.)  Each octant has
 second priority in two of these regions (adjacent) and third priority in
another two (non-adjacent).

Whereas for $P_z=1$ the secondary and tertiary quadrants finished a sector
before the final quadrant moved in, for $P_z>2$ we see six octants at play in a
sector.  The three octants with directions opposite $R$, $B$ and $G$, which we
will call $\bar R$, $\bar B$ and $\bar G$, arrive before the first three are
finished.  The boundary planes for each of these with the $RBG$ octants that are
not its inverse are the sector boundaries.  The boundary planes with their
opposites are perpendicular to the problem domain's diagonals; each of these
carves up two of the six regions where the second octant of $RBG$ was
unchallenged.

Since $D(O) + D(\bar O) = \mathrm{constant}$, the top four octants for a
priority region are reversed for the final four.  For example, a region with
$(P, R, G, \bar B,...)$ must in fact have the priorities $(P, R, G, \bar B, B,
\bar G, \bar R, \bar P)$.  {(We will take this region as our example in
the description that follows.)}  These {divisions give} us twelve
distinct priority regions per sector.


\subsubsection{First priority octants}

Things begin much as before, with eight corner processes initiating waves of
tasks in their primary octants.  The stage counts are the same:
\begin{equation}
\mu(m, i, j, k) = s + m  \; , \text{ where $s$ = stage index and $m$ = angleset index.}
\end{equation}
The waves propagate to the central processes in $X+Y+Z-2$ steps, as in Eq.\
(\ref{eqn:nfill}).  The central processes receive all primary-octant tasks in
smooth succession, and they begin to satisfy their neighbors' dependencies.


\subsubsection{Second, third and fourth priority octants}

The second-, third- and fourth-priority octants ($R$, $G$ and $B$) collide with the
first-priority octant (P) at three of the sector's corners, and the standing
collision fronts spread across the sector boundaries.  Once the tail of the P
wave passes, the $R$, $G$ and $B$ waves begin propagating inward from those points.
They initially propagate smoothly, each with delay of $M-1$ stages.

These three octants collide with each other at lines, and their collision fronts
spread from there.  They all reach the central process at the same stage, $X +
Y + Z + M - 2$, and the central process begins its second-priority tasks.
In the $P_z=1$ case, everything holds static while the central process
executes its second-priority tasks, but in the $P_z>2$ case, the $R$, $G$ and $B$ octants
move into their tertiary regions (as in the rightmost image in Fig.~\ref{fig:rgb}) during this phase. 
They suffer a
delay of $M$ at these interfaces, and then continue propagating in a sort of
rotation around the central process.  Once the central process is ready, it
moves on to its third- and fourth-priority octant, which have been ready for it
since it began its second-priority tasks.


\subsubsection{Fifth, sixth and seventh priority octants}

As can be seen from sector symmetry, the next octants have been ready to enter
the sector since the $R$, $G$ and $B$ waves collided.  However, the entry
processes had $2M$ tasks available with higher priority.  Once these are
done (e.g., once the final $R$ and $G$ tasks are propagating along the RG
boundary), the next octant (e.g., $\bar B$) enters the sector.  It has been
delayed $3M-2$ stages in total, and propagates as $\mu = s + m + 3M - 2$.

Each octant collides with its opposite at an entire plane, where each side
finishes its priority.  When they switch sides, they continue with a planar
wavefront at a delay of $M$ (or $M-1$ for the winner of the tie-breaker).
As the waves continue to intersect, they continue to delay but not disrupt each
other, and they all sort of pivot around the central process.

While there are many collisions and priority regions in a sector, the central
process is never in danger of running out of available work.  Thus, the second 
condition for optimal scheduling is met.


\subsubsection{Final octants}

Symmetry assures that each central process's dependencies for its final octant
are met before it finishes the other octants.  The previous octants have already
propagated well past the central process, fulfilling dependencies as they
went.  The final octants meet no competition on their way to the problem
boundary.

Throughout this choreography, the fundamental requirements of an optimal scheduling
algorithm are met.  The central processes get their work as early as possible,
they stay busy until they are done, and their final tasks propagate freely to the corners 
of the problem domain, marking the end of the eight-octant boundary-to-boundary sweep.


\subsection{$P_z>2$ Decomposition with $\omega_z>1$}

The optimal scheduling strategies described for particular cases of partitioning 
and aggregation in previous subsections also apply to the remaining case, in which $P_z>2$ 
and $\omega_z>1$.  This is a merger of the ``hybrid'' ($P_z=2, \omega_z \ge 1$) 
and ``volumetric'' ($P_z>2, \omega_z=1$) cases.

In the case of $P_z=2$ and $\omega_z \ge 1$, the central process receives its first task after 
$X + Y -2$ stages, and there are $\omega_m \omega_z \omega_g$ work stages.  
In the case of  $P_z>2$ and $\omega_z=1$, the central process receives 
its first task after $X+Y+Z-3$ stages, and there are $\omega_m \omega_g$ work stages. 
In previous subsections we showed that the 
for these cases, the scheduling algorithms described herein (such the ``depth-of-graph'' 
algorithm) execute full eight-octant sweeps in the minimum possible number of stages.

For the more general case of $P_z>2$ and $\omega_z>1$, it remains true that the 
scheduling algorithms described herein execute the sweep in the minimum number 
of stages.  Showing this requires the same kinds of arguments used for previous cases.

One might ask whether it ever makes sense to have $\omega_z>1$ when $P_z > 2$.  
Sometimes it does.  Recall that an important ingredient in parallel efficiency is the ratio 
of idle-stage count to working-stage count:
\begin{align}
  \text{ratio} & = \frac{P_x + \delta_x + P_y + \delta_y -4+ \omega_z(P_z+\delta_z-2)}
    {\omega_m\omega_g\omega_z}
\end{align}
If all $P_u>2$, which is the case under discussion, then we see that increasing $\omega_z$ 
while holding all other variables constant \emph{decreases} the idle-to-working ratio.  Of 
course, it also increases the communication-to-working ratio by making more, smaller 
tasks.  But in practice we find that the optimal partitioning and aggregation for large problems, 
taking everything into account, includes $P_z>2$ 
and $\omega_z>1$.


\subsection{Reflecting Boundaries}

Our decomposition and scheduling algorithms have mirror symmetry across a problem's
$x$-$y$, $x$-$z$, and $y$-$z$ planes.  This leads to the desirable property that problems
with reflecting boundaries will execute exactly like their full-domain counterparts.
That is, from the perspective of the processes
assigned to a given portion of the problem, it makes no difference if incoming angular fluxes on a central
mirror plane of a symmetric problem come from processes in the neighboring portion of the full spatial domain or from
reflection of outgoing angular fluxes---in either case the algorithm ensures that those tasks
are available when it is time to execute them.  That is, for example, full-domain execution of a symmetric
3D problem with $P$ processes proceeds exactly like execution of one eighth of the problem
with $P/8$ processes and three reflecting boundaries.

%% file: s4-optimal.tex
\section{OPTIMAL SWEEPS}
\label{sec:opt_sweeps}

Here we describe how we have used our optimal {\em scheduling} algorithm to
generate an optimal {\em sweep} algorithm.  Given an optimal schedule we know
exactly how many stages a complete sweep will take, and thus can estimate the
parallel efficiency of a sweep with such a schedule:
\begin{eqnarray} \label{eq:opteb}
\epsilon_{opt} = \frac{1}{
  \left[1 + \frac{P_x + \delta_x + P_y + \delta_y -4+ \omega_z(P_z+\delta_z-2)}
    {\omega_m\omega_g\omega_z} \right]
  \left[1 + \frac{T_\mathrm{comm}}{T_\mathrm{task}}\right]}.
\end{eqnarray}

Given Eq.\ (\ref{eq:opteb}), we can choose the \{$P_x, P_y, P_z, \omega_m, 
\omega_g, \omega_z$\} that
maximize efficiency and thus minimize total sweep time.  This optimization over the
\{$P_u$\} and \{$\omega_j$\}, coupled with the scheduling algorithm that executes the
sweep in $N_{stages}^{min}$ stages, yields what we call an {\em optimal sweep}
algorithm. 

The denominator of the efficiency expression is the product of two terms, and
optimization means minimizing this product.  Several observations are in order.
First, aggregation into a larger number of smaller tasks causes the first term to 
decrease (because
$\omega_m\omega_g\omega_z$ is the number of tasks) and the second term to
increase (because $T_{task}$ shrinks while the latency portion of $T_{comm}$
remains fixed).  Thus, for a given \{$P_u$\} and given problem size there will
be some set of aggregation parameters that minimize the product.  

Second, the
term $(P_z+\delta_z-2)$ vanishes when $P_z=1$ {\it or} $2$, leading to the
benefit of our ``hybrid'' partitioning discussed above:  If we change from $P_z=1$
to $P_z=2$ and keep processor count and task size constant, the first term decreases
(because $P_x +P_y$
decreases) and the second stays about the same (because
$T_{task}$ stays the same).  

Third, if we use  $P_x \approx P_y \approx P_z$ and $\omega_z=\omega_y=\omega_x = 1$ (the usual ``volumetric'' decomposition 
strategy), then $P_x+P_y+P_z$ grows as $P^{1/3}$ instead of the $P^{1/2}$ that 
occurs when $P_z$ is fixed equal to 1 or 2.  This hints that for very high processor counts 
a volumetric decomposition might be best.

It is interesting to compare $\epsilon_{\text{KBA}}$ 
(which uses $P_z=1$ and sweeps two octants at a time) to 
$\epsilon_{opt,hyb}$ (which uses $P_z=2$ and sweeps all eight octants simultaneously),
especially in the limit of large $P$ (which allows us to ignore the $\delta_u$
and the numbers $2$ and $4$ that appear in the equations).  In the large-$P$
limit, with $P_x+P_y \approx P^{1/2}+P^{1/2}$, Eq.\ (\ref{eqn:kbae}) becomes
\begin{eqnarray} \label{eqn:kbae2}
\epsilon_{\text{KBA}} \xrightarrow{\text{large } P}
  \frac{1}{\left[1+\frac{4(2P^{1/2})}{\omega_m\omega_z} \right]
           \left[1+\frac{T_{comm}}{T_{task}}\right]} 
\end{eqnarray}
Now consider $\epsilon_{opt,hyb}$ with $P_z=2$ and $P_x+P_y \approx 2(P/2)^{1/2} = \sqrt{2}P^{1/2}$.
For comparison we aggregate to the same number of tasks as in KBA (which is
likely sub-optimal for hybrid), so $\omega_g=1$ and $\omega_m$ is the same as in KBA.  The result is
\begin{eqnarray} \label{eqn:opte2}
\epsilon_{opt,hyb} \xrightarrow{\text{large P}}
\frac{1}{\left[1+\frac{\sqrt{2}P^{1/2}}{\omega_m\omega_z} \right]
         \left[1+\frac{T_{comm}}{T_{task}}\right]}.
\end{eqnarray}

An interesting question is how many more processors the hybrid partitioning with optimal scheduling can use
with the same efficiency as what we have called ``basic'' KBA.  The answer comes from setting $4(2 P_{KBA}^{1/2}) = \sqrt{2}P_{opt,hyb}^{1/2}$, which yields the result $P_{opt,hyb}/P_{KBA} = 32$.
For example, even without
optimizing the \{$\omega_j$\}, our 8-octant scheduling algorithm with $P_z=2$ yields the same efficiency
on 128k cores as ``basic'' KBA on 4k cores.  Optimizing the \{$\omega_j$\} can
improve this even further.  The improvement stems from launching all octants 
simultaneously, which significantly reduces process idle time, and managing the ``collisions'' of the multiple sweep fronts in a way that does not add extra stages.  The cost is that 
more storage is required during the sweep, because the angular fluxes on all of the 
sweep fronts must be stored at the same time.

Our simplest performance model is Eq.\ (\ref{eq:opteb}) with the following definitions:
\begin{equation}\label{eqn:tcomm}
  T_{comm} = M_L \times 3 \times T_{latency} + T_{byte} N_{bytes}
\end{equation}
\begin{equation}\label{eqn:ttask}
  T_{task} = T_{wu} + A_x A_y A_z \left(T_{cell}
            + A_m \left[ T_m + A_g T_{g} \right] \right)
\end{equation}
where 
\begin{align*}
  T_{latency} &= \text{message latency time,} \\
  T_{byte} & =  \text{additional time to send one byte of message,} \\
  N_{bytes} & = \text{total bytes a processor must send to downstream neighbors
   at each stage,} \\
  T_{wu} & = \text{time per task outside of comms and outside loop over cells in
    cellset} \\
  T_{cell} & = \text{time per task per spatial cell, outside loop over
    directions in angleset} \\
  T_m & = \text{time per task per spatial cell per direction, outside loop over
    groups in groupset} \\
  T_{g} & = \text{time in inner (group in groupset) loop to compute a single
    cell, direction, and group.}
\end{align*}

$N_{bytes}$ is calculated based on the aggregation
and spatial discretization scheme; the other parameters are obtained through
testing.  We use the parameter $M_L$ to explore performance as a function of
increased or decreased latency.  The factor of 3 in the latency term is because 
processors typically must send three messages at each stage of the sweep.  
If we find that a high value of $M_L$ is needed
for our model to match our computational results, then we look for things to
improve in our code implementation.

We have implemented in our PDT code an ``auto" partitioning and aggregation
option.  When this option is engaged, the code uses empirically determined
numbers for $T_{latency}$, $T_{byte}$, $T_{wu}$, $T_{cell}$, $T_m$, and $T_{g}$ for the
given machine.  Then for the given problem size it searches for the combination
of $\{P_u\}$ and $\{A_j\}$ that minimizes the estimated solution time.  This relieves
users of the burden of choosing these parameters and ensures that efficient choices are made.  In the numerical
results shown in the following section we did not employ this option, because we
were exploring variations in performance as a function of aggregation parameters
and thus wanted to control them.  However, we often use this option when we use
the PDT code to solve practical problems.

Angle aggregation carries complexities that group and cell aggregation do not.  We
mentioned in Sec.~\ref{sec:intro} that all directions in an angleset must belong
to the same octant, for otherwise they would need to start on different corners of 
the spatial domain.  In PDT, the directions in an angleset
must all share a sweep ordering, because the loop over directions in an angleset 
is inside the loop over cells.  If the grid has only brick-shaped cells, then all directions 
in a given octant have the same cell-to-cell dependencies.  
In a completely unstructured grid, though, the cell-to-cell 
sweep ordering (dependence graph) that respects 
all upstream dependencies can be different for each quadrature direction.  
In the current version of PDT, a sweep that respects all dependencies 
would in such a situation
require a different angleset for each quadrature direction.  An alternative is to 
relax the strict enforcement of dependencies, using previous-iteration information 
for angular fluxes from upstream cells that have not yet been calculated during 
the sweep.  For example, if cell $i$ is calculated before cell $j$ (because $i$ is 
the upstream cell for most of the directions in the angleset), but for some directions 
in the angleset cell $j$ is upstream of cell $i$, then for those particular directions 
the $j$-to-$i$ angular flux would come from the previous iteration.  If this happens 
extensively, it can increase iteration counts.  The ideal approach will be problem-dependent.

We mentioned in
Sec.~\ref{sec:intro} that 3D grids of polygonal-prism cells (polygons in a plane, 
extruded into the third dimension) can offer advantages for sweeps, relative 
to fully unstructured polyhedral grids.  This is most pronounced when prismatic-cell 
grids are used with ``product'' quadrature sets, which have multiple directions that have different polar angles 
(angles relative to the axis of prismatic extrusion) but the same azimuthal angle (angle 
in the plane of the polygons).  All directions with the same azimuthal angle in the 
same octant have the same cell-to-cell sweep ordering on prismatic grids, 
which allows them to be aggregated without resorting to previous-iteration information.   We typically take advantage of this in problems that lend themselves to prismatic grids, including the 3D nuclear-reactor problems illustrated in the next section.

It takes much more than a good parallel algorithm to achieve the scaling 
results that we present in the following section.
Implementation details are important for any code that attempts to
scale up to and beyond $10^6$ parallel processes.  Our PDT results are due in no small part to
the STAPL library, on which the PDT code is built.  STAPL provides parallel data
containers, handles all communication, and much more.  See
\cite{stapl-1,stapl-2,stapl-3,stapl-4,stapl-14,stapl-15a,stapl-15b,stapl-16,stapl-17} and \cite{stapl-git} for more details.  We also present results from LLNL's ARDRA code, which has benefited from LLNL's long experience in efficient utilization of the world's fastest computers.

%% file: s5-results.tex
\section{COMPUTATIONAL RESULTS} 
\label{sec:results}

In this section we present results from a series of test problems that demonstrate 
how sweep times at high core counts compare to those at low core counts when 
our optimal sweep algorithm is used.  We begin with 
simple brick-cell weak-scaling suites and then turn to a weak-scaling suite with 
spatial grids that resolve geometric features in a pressurized-water 
nuclear reactor.  

\subsection{Regular Brick-Cell Grids, DFEM Spatial Discretization}

We begin with suites of brick-cell test problems in which as $P$ grows, 
the size of the problem domain
increases while cell size, cross sections, and number of cells per parallel process 
are unchanged.  The simplest version has only one energy group, 80 directions 
(10 per octant), and 4096 cells per core.  We employ the PWLD spatial discretization \cite{pwld-a,pwld-b}, which has 8
unknowns per brick-shaped cell (one for each vertex).  We will see later that 
problems with more groups and angles exhibit higher
parallel efficiencies.

We studied weak scaling, holding constant the number of unknowns per processing
unit.  We ran this series from $P=8$ to $P=384$k $=384\times 1024 = 393,216$ 
cores, with the
depth-of-graph and push-to-central scheduling algorithms.  With an earlier
version of our code we also tested a non-optimal scheduling algorithm that 
simply executes tasks in the order
they become ready, from $P=8$ to $P=128$k =131,072 cores.  The problems were run
on the Vulcan computer at LLNL, an IBM BG/Q architecture with 16 cores per node.

All results are for $P_z=2$, and efficiencies are based on solve times
normalized to a $P=8$ run.  Times do not include setup but {do} include
communications and convergence testing.

\begin{figure}[h!]
\centering
\includegraphics[width = 0.75\linewidth]{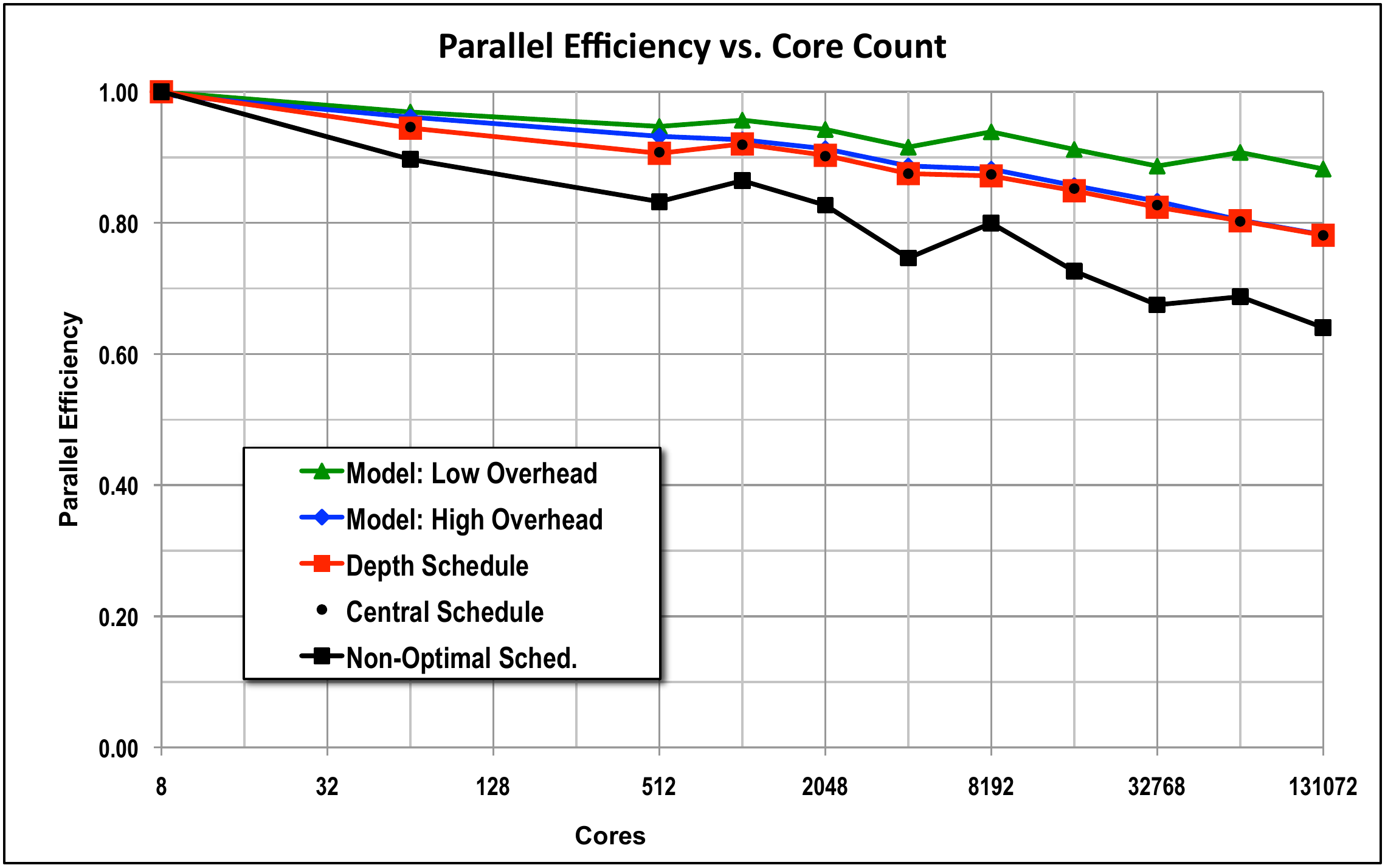}
\caption{Weak scaling results from three different
scheduling algorithms on an IBM BG/Q computer.  ``Efficiency'' is 8-core execution time divided by the $P$-core time, with unknowns/core held constant.  Red squares and black dots show observed results from the PDT code with two different optimal scheduling algorithms, depth-of-graph and push-to-center, respectively.  They are identical, as predicted by theory.  Black squares represent PDT results from a first-come-first-serve schedule algorithm, which is sub-optimal.  Green triangles are predictions of the idealized performance model of Eq.~(\ref{eq:opteb}).  Blue diamonds (mostly hidden by PDT optimal-schedule results) are from the same model with the latency multiplier $M_L=11$.}
\label{fig:scaling_pttl}
\end{figure}

Figure~\ref{fig:scaling_pttl} shows results  three different
scheduling algorithms:  the depth-of-graph and push-to-central optimal schedules
and the (non-optimal) first-arrival schedule.  We have compared the observed stage
counts against the minimum-possible stage counts described previously in this
paper, and in every case they agree exactly for both of the scheduling
algorithms that our theory claims are optimal.  The figure indicates that the
non-optimal schedule does not perform as well as the optimal schedules, but it
degrades surprisingly slowly.  We see that sweeps executed with optimal
schedules perform very efficiently out to large core counts, even with a
modest-sized problem (only one group and 80 directions).

\begin{figure}[h!]
\centering
\includegraphics[width= 0.75\linewidth]{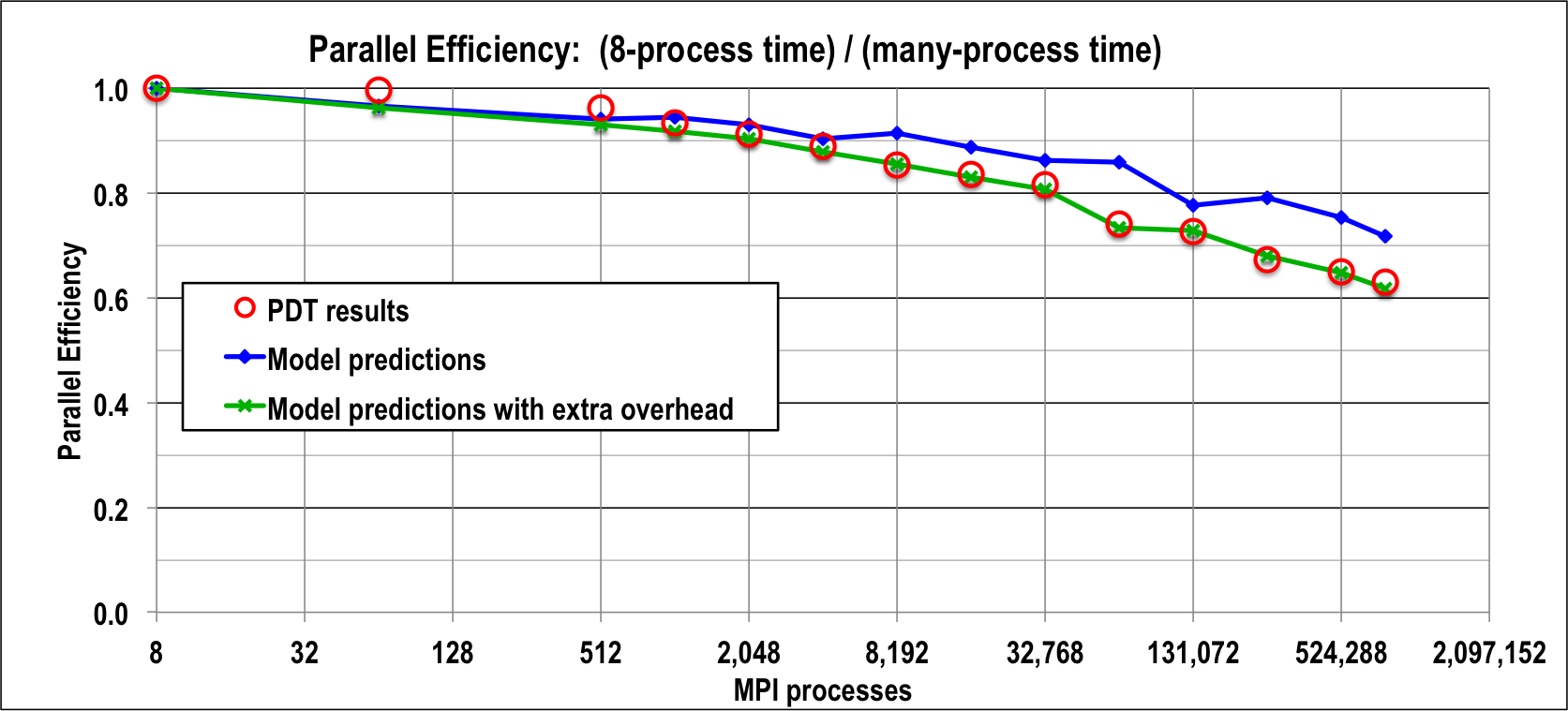}
\caption{Weak scaling results from optimal sweep algorithm on IBM BG/Q computer.  ``Efficiency'' is 8-core execution time divided by the $P$-core time, with unknowns/core constant and one MPI process per core.  Red circles show results from PDT on one-group problems with 10 directions per octant, 4096 brick-shaped spatial cells per core, and one MPI process per core.  The line with diamond markers is efficiency predicted by the performance model of Eq.~(\ref{eq:opteb}) with $M_L=1$.  The line with $\mathbf{\times}$ is from the model with $M_L=11$.  The highest core count shown is the entire Mira machine at Argonne National Lab: 768$\times$1024 cores.}
\label{fig:mira1g}
\end{figure}

Figure~\ref{fig:mira1g} provides results out to 768k cores for a three-group
version of our test problem using the
push-to-center optimal scheduling algorithm.  Even though the test problem 
had only three energy groups and only 10 quadrature directions per octant, 
the code achieved more than 60\% parallel efficiency when scaling up from 
8 to 786,432 cores.  That is, the optimal sweep algorithm loses less than 
40\% efficiency when scaling up by a factor of 96k on this small test problem.

Figure~\ref{fig:mira1g} also shows efficiency predictions of our performance model for
two different overhead burdens.  The ``low-overhead" plot used $M_L=1$ in Eq.\
(\ref{eqn:tcomm}), which is what we would hope to achieve in a nearly perfect
implementation of our algorithms in our code.  In this case the only overhead
would be actual message-passing latency.  The ``high-overhead" plot used
$M_L=11$, and it agrees closely with our observed performance.  This suggests
that there is per-task overhead in our code implementation that we should be
able to reduce.  We are working on this.

\begin{figure}[h!]
\centering
\includegraphics[width= 0.75\linewidth]{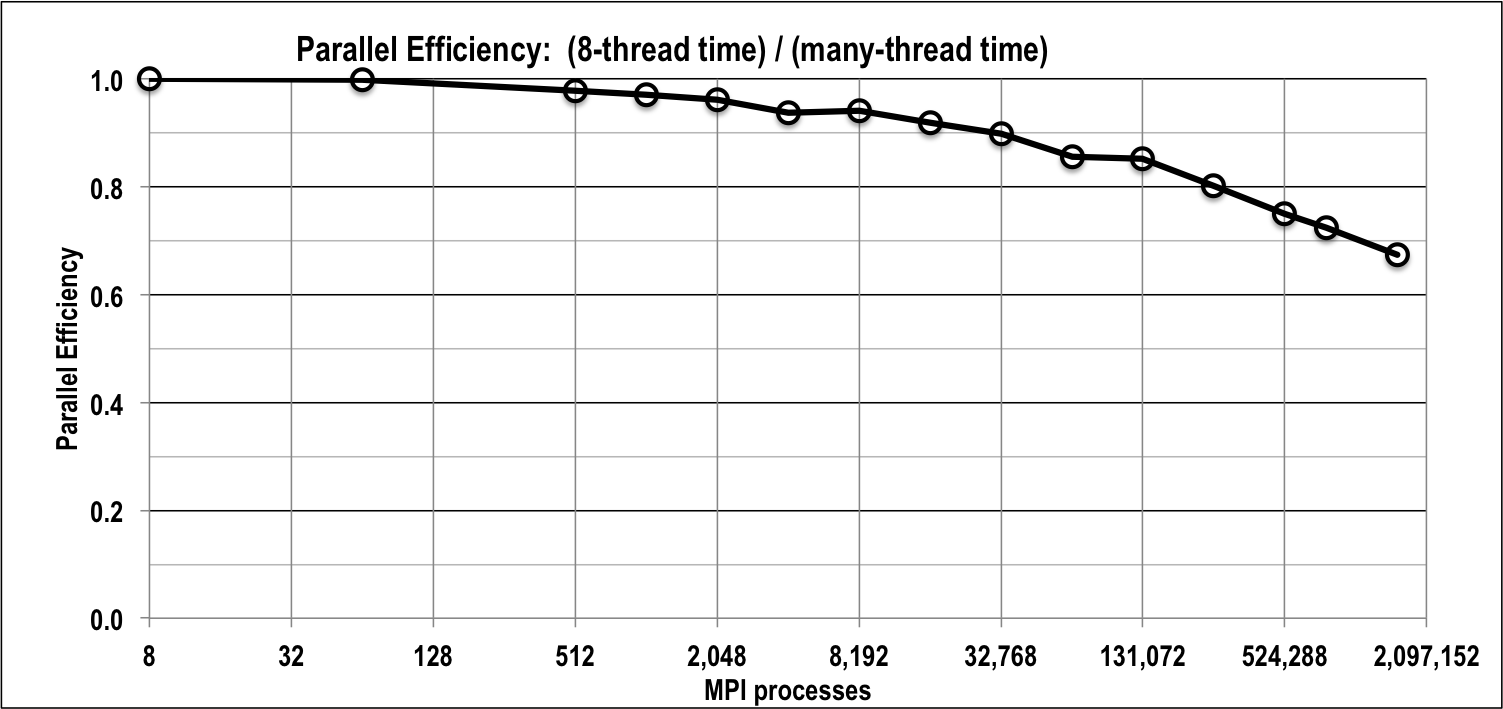}
\caption{Weak scaling results from optimal sweep algorithm on IBM BG/Q computer.  ``Efficiency'' is the time to execute a sweep on a given number of processes divided by the time to execute a sweep on 8 processes, with work per process held constant.  Results are from the PDT code on a series of 3-group problems with 10 directions per octant, 4096 brick-shaped spatial cells per core, and two MPI processes per core.  The highest process count shown is 2$\times$768$\times$1024 = 1,572,864 MPI processes.}
\label{fig:mira3g2tpc}
\end{figure}

Continuing to push to higher parallelism, we executed a weak scaling study out to $\approx 1.6$ million ($3 \times 2^{19}$) parallel threads by ``overloading'' each of the $768 \times 1024$ cores with 2 MPI processes.  Our test problem is as before---4096 brick cells per thread,  10 directions per octant, and three energy groups.  As Fig.~\ref{fig:mira3g2tpc} shows, the optimal sweep algorithm in PDT achieved 67\% parallel efficiency when scaling from 8 threads to approximately 1.6 million threads---it loses only 33\% efficiency when scaling up by a factor of 192k when there are three energy groups.  Scaling improves further with more groups or more directions, because the work-to-communication ratio improves.
%

\subsection{Regular Brick-Cell Grids, Diamond Differencing Spatial Discretization}

ARDRA is a research code developed at LLNL to study parallel discrete ordinates transport.  The code applies a general framework to domain decompose the angle, energy and spatial unknowns among available parallel processes. Typically, problems run with ARDRA are decomposed only in space (volumetrically) and energy.  Spatial overloading is not currently supported, so one cellset equals one process's subdomain. In addition, ARDRA does not aggregate directions, which means a single direction per angleset.  ARDRA's default spatial discretization is diamond differencing, with only one spatial unknown and only a few operations required to solve each cell.  

The model of the time to completion for this algorithm is:
\begin{align}
  T_{solve} & = G \; N_{stages} \left( T_{task} + T_{comm} \right) + T_{RHS} = N_{stages} \; \tilde T + T_{RHS} \; ,
\end{align}
where $G$ = number of groups and $T_{RHS}$ is the time to calculate the scattering source.  Note that this time includes both sweep time and source-building time.  With spatial-only decomposition, the source-building operation does not require communication among processes, and thus it is somewhat easier to scale well on total solve time than on sweeps alone.  The corresponding efficiency model is:
\begin{align}
  \epsilon & = \frac{T_{ref}}{T_p} = \frac{N_{stages}^{ref} \; \tilde T + T_{RHS}}{N_{stages}^p \; \tilde T + T_{RHS}} 
   = \frac{N_{stages}^{ref} \; + T_{RHS}/\tilde T}{N_{stages}^p + T_{RHS}/ \tilde T}
   \label{eq:ardraeff}
\end{align}

The Ardra scaling results shown here are based on the Jezebel criticality experiment.  We ran this problem in 3D with all vacuum boundary conditions, 48 energy groups, and three level-symmetric quadrature sets:  S$_8$ (80 directions), S$_{12}$ (168), and S$_{16}$ (288).  We performed two weak scaling studies:  one with spatial parallelism only, and the second with a mixture of energy and spatial parallelism.  We ran standard power iteration for $k$-effective, stopping the run at 11 iterations, which was adequate for collecting timing statistics.  Both of our weak scaling studies start with one node of Sequoia (an IMB BG/Q machine), using 16 MPI ranks, with 1 rank per CPU core.  

Both studies have an initial $48 \times 24 \times 24$ spatial mesh, but decompose the problem differently across the 16 ranks. In our first weak scaling study we decompose the problem into $12 \times 12 \times 12 = 1792$ cells per rank, with the resulting spatial decomposition on $N_{nodes}$ Sequoia nodes of $P_x = 4 N_{nodes}, P_y = 2 N_{nodes}$, and $P_z = 2 N_{nodes}$. Our second study uses 16-way on-node energy decomposition, with each rank having $48 \times 24 \times 24 = 16 \times 1792$ spatial cells but only 3 energy groups.  Weak scaling is achieved by increasing the number of spatial cells proportional to increasing processor count.  

\begin{figure}[h!]
\centering
\includegraphics[width= 0.65\linewidth]{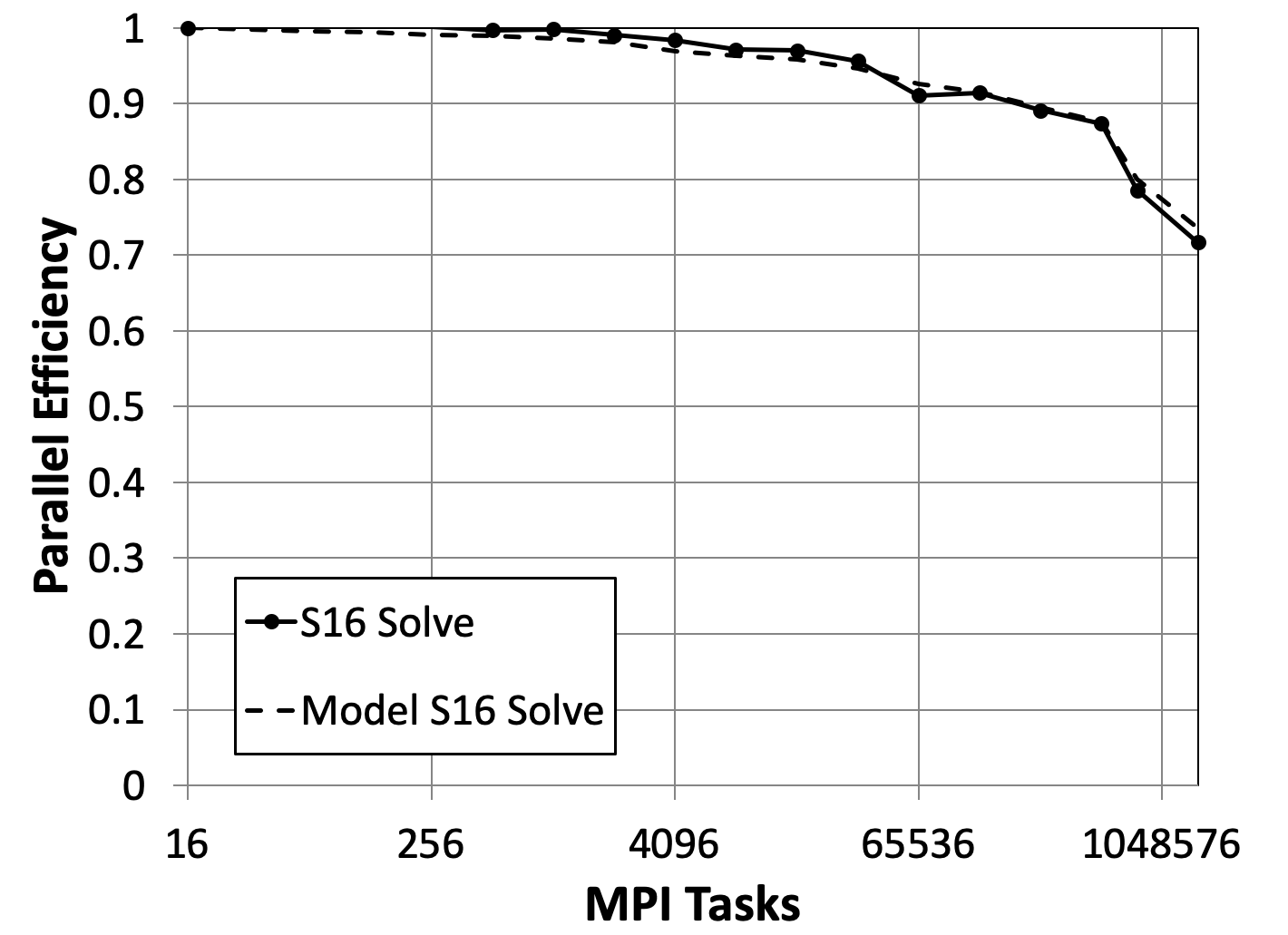}
\caption{Weak scaling results:  total solution time (sweep plus other) from Ardra, with combined spatial and energy parallel partitioning and the S$_{16}$ level-symmetric quadrature set, on the Sequoia IBM BG/Q computer.   The highest process count shown is 1,572,864 MPI processes (one per core, entire machine).}
\label{fig:ardras16}
\end{figure}

Ardra's largest run was at Sequoias's full scale, which is 37.5 trillion unknowns using 1,572,864 MPI ranks.  With the  achieved 71\% parallel efficiency for total solution time, when using both energy and spatial parallel decomposition and the S$_{16}$ quadrature set, as shown in Fig.~\ref{fig:ardras16}.  The figure also shows excellent agreement between observed results and the performance model of Eq.~(\ref{eq:ardraeff}).

\begin{figure}[h!]
\centering
\includegraphics[width= 0.65\linewidth]{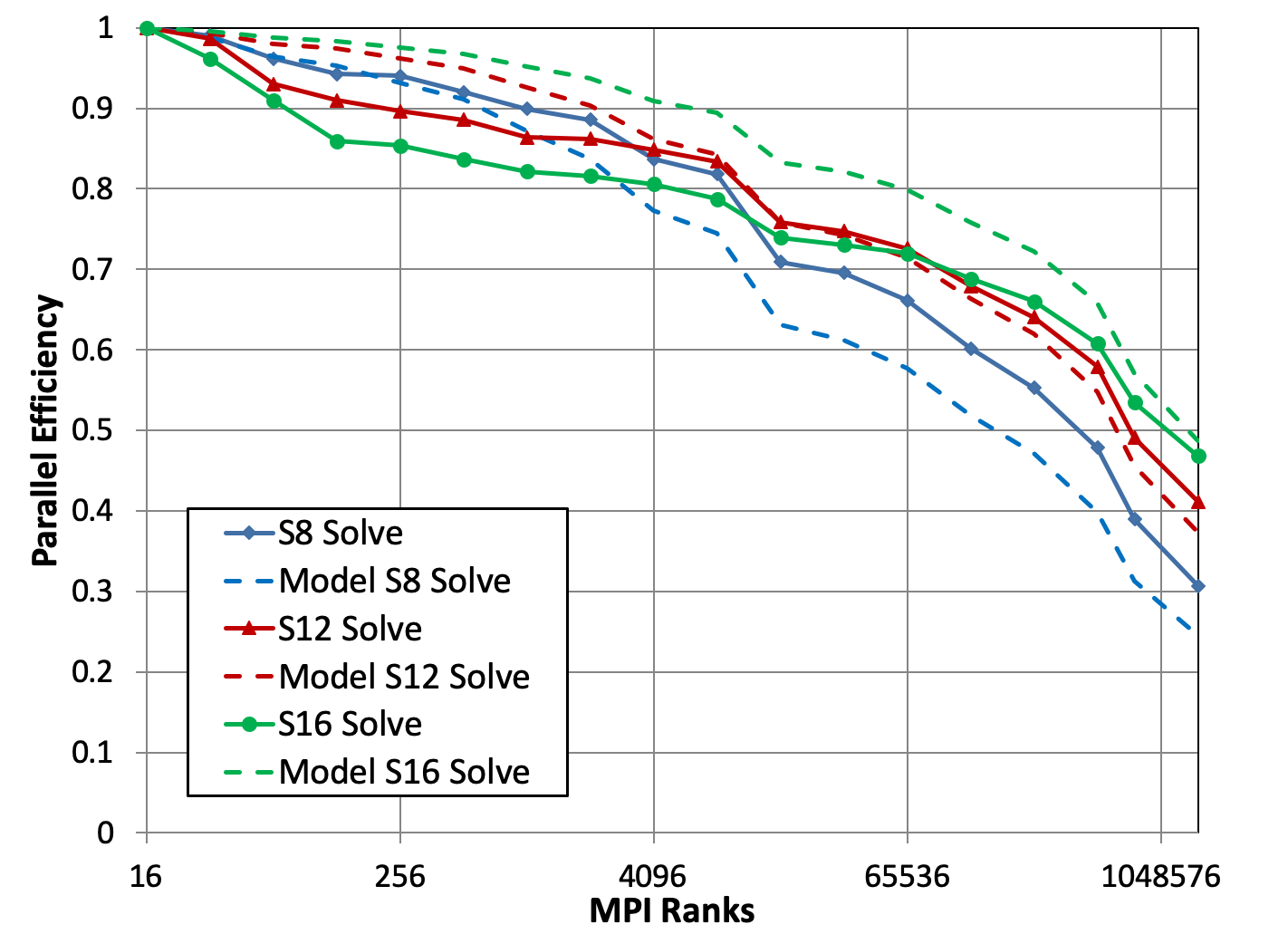}
\caption{Weak scaling results from Ardra solution on the Sequoia IBM BG/Q computer.   The highest process count shown 1,572,864 MPI processes (one per core, entire machine).}
\label{fig:ardrasolvesp}
\end{figure}

\begin{figure}[h!]
\centering
\includegraphics[width= 0.65\linewidth]{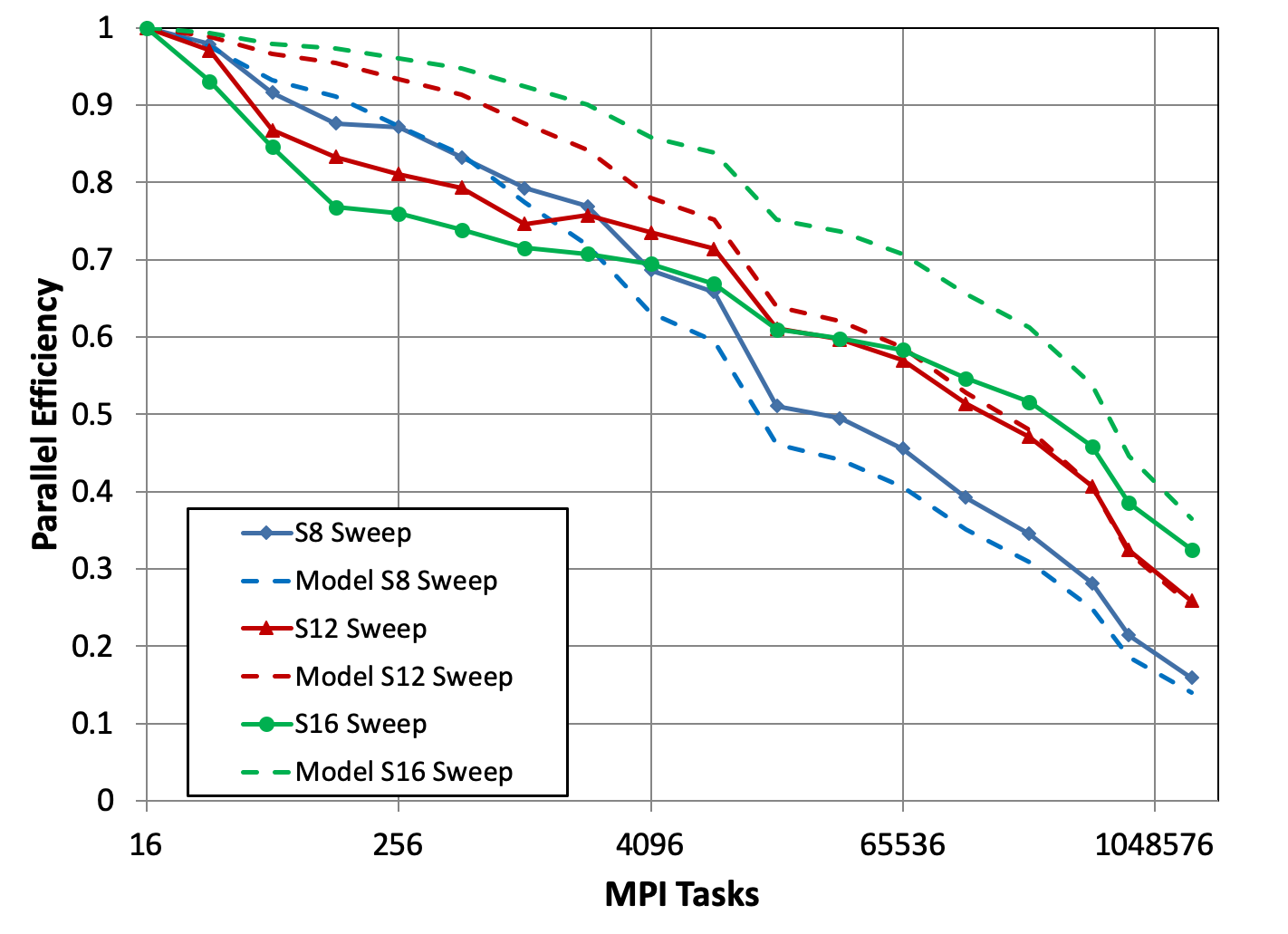}
\caption{Weak scaling results from Ardra solution on the Sequoia IBM BG/Q computer.   The highest process count shown 1,572,864 MPI processes (one per core, entire machine).}
\label{fig:ardrasweepsp}
\end{figure}

Figures \ref{fig:ardrasolvesp} and \ref{fig:ardrasweepsp} give efficiency results for all three quadrature sets on the test suite that used spatial-only decomposition.  We offer several observations.  First, the performance model is not perfect but does capture the trends observed in the ARDRA results.  Second, total solve time scales much better than sweep-only time.  Third, scaling improves substantially with increasing number of quadrature directions.  This is easy to understand given that ARDRA is using only a single cellset per process, which means directions are the only means available for pipelining the work and getting the central processes busy.  Fourth, comparison of the S$_{16}$ results from the figures shows that for this problem with this code, parallelizing across energy groups is a substantial win, moving parallel efficiency from just under 50\% to just over 70\% at a core count of 1.5M.


\subsection{Reflecting Boundaries}

Reflecting boundaries introduce \emph{dependencies} among octants of directions, and these dependencies hamper parallel performance.  For example, in a problem with two reflecting boundaries that are orthogonal to each other (i.e., not opposing), only two octants of directions (not all eight) can be launched in parallel at the beginning of the sweep.  It turns to be straightforward to quantify the performance of our optimal sweeps with reflecting boundaries in terms of the performance without reflecting boundaries.

Previously we mentioned that in our algorithm, at a reflecting boundary a processor feeds itself incident fluxes (by reflecting them from outgoing fluxes) at the same stages in the sweep at which a neighboring processor would feed them if the full problem domain were being run with twice as many processors.  Consider a problem with reflective symmetry at $x=0$ and at $y=0$.  If we run this problem with $4P$ processors on the full domain $x \in (-a,a) \times y \in (-b,b)$, we therefore expect essentially the same performance as if we run with $P$ processors on the reflected quarter domain $x \in (0,a) \times y \in (0,b)$.  The difference:  in the $4P$-processor full-domain case communication is required to feed the angular fluxes, whereas in the $P$-processor quarter-domain case a calculation is done to perform the reflection.  In our experience the differences are negligible (a few percent), with the full problem sometimes slightly faster (with 4$P$ processors) and the quarter problem sometimes slightly faster (with $P$ processors).

It follows that the performance of our algorithm with $P$ processors on a problem with two reflecting boundaries is roughly the same as the performance with $4P$ processors on problems without reflecting boundaries, or with $8P$ processors on problems with three (mutually orthogonal) reflecting boundaries.  This quantifies the sweep-efficiency penalty introduced by reflecting boundaries:  if there are $n$ reflecting boundaries, then efficiency with $P$ processors is only what would be expected from $P \times 2^n$ processors on the full problem.  This allows us to demonstrate how our sweep methodology would perform on up to 8 times as many processors as are actually available.  

In the following section we test our sweeps on polygonal-prism grids that accurately represent interesting nuclear-reactor problems.  In these problems there is often reflective symmetry on two orthogonal boundaries; thus, they present an opportunity to test how our sweeps would perform out to four times as many cores as are actually available to us.

\subsection{Polygonal-Prism Grids}

We turn now to spatial grids that can represent complicated geometric structures with high fidelity.  In particular, we consider grids composed of right polygonal prisms, which are well suited to representing structures that have arbitrary complexity in two dimensions but some regularity in the third dimension.  Nuclear reactors with cylindrical fuel pins are a good example and are the basis for the test problems we consider next.

\begin{figure}[h!]
\centering
\includegraphics[width= 0.4\linewidth]{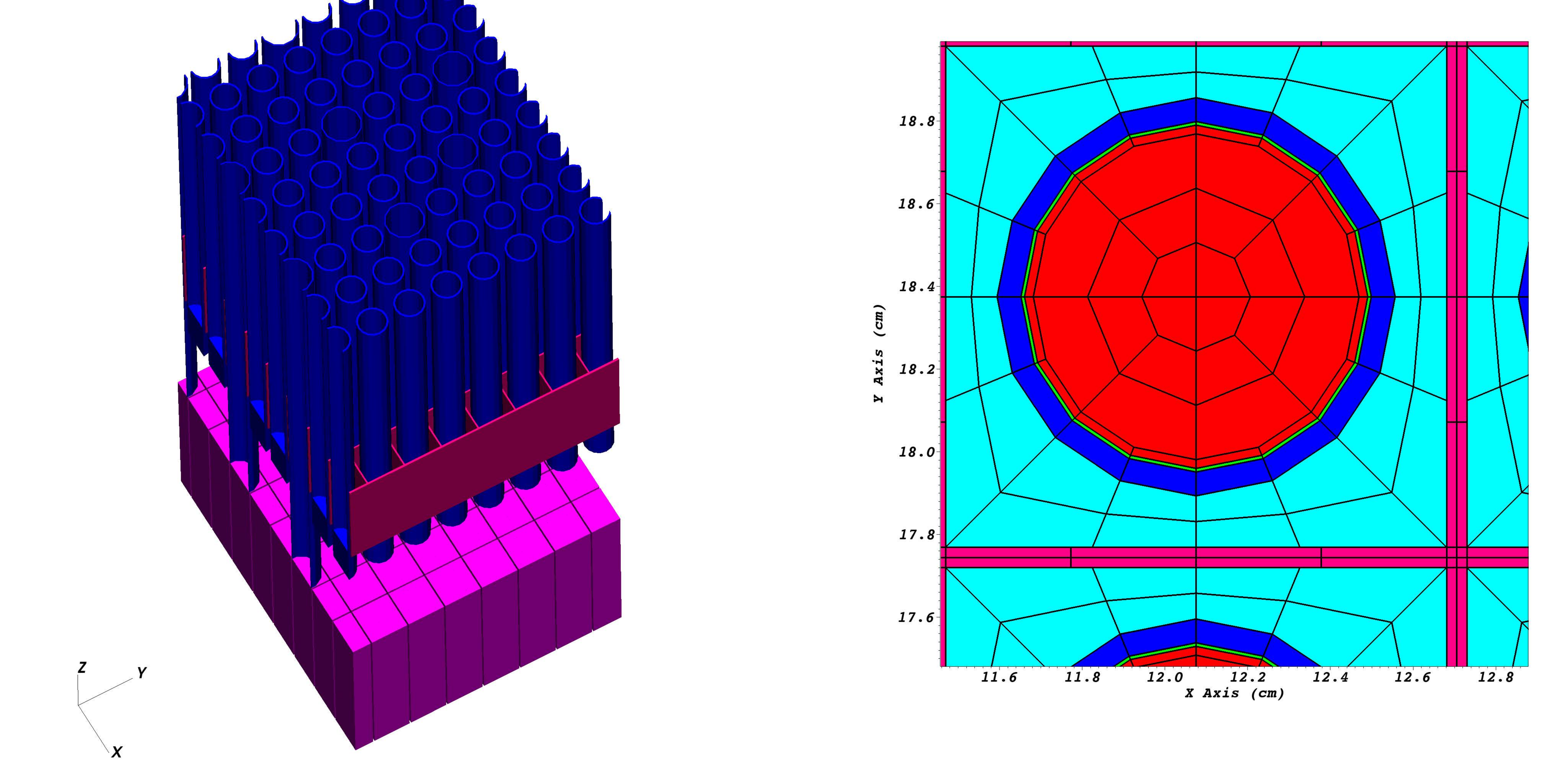} \qquad
\includegraphics[width= 0.45\linewidth]{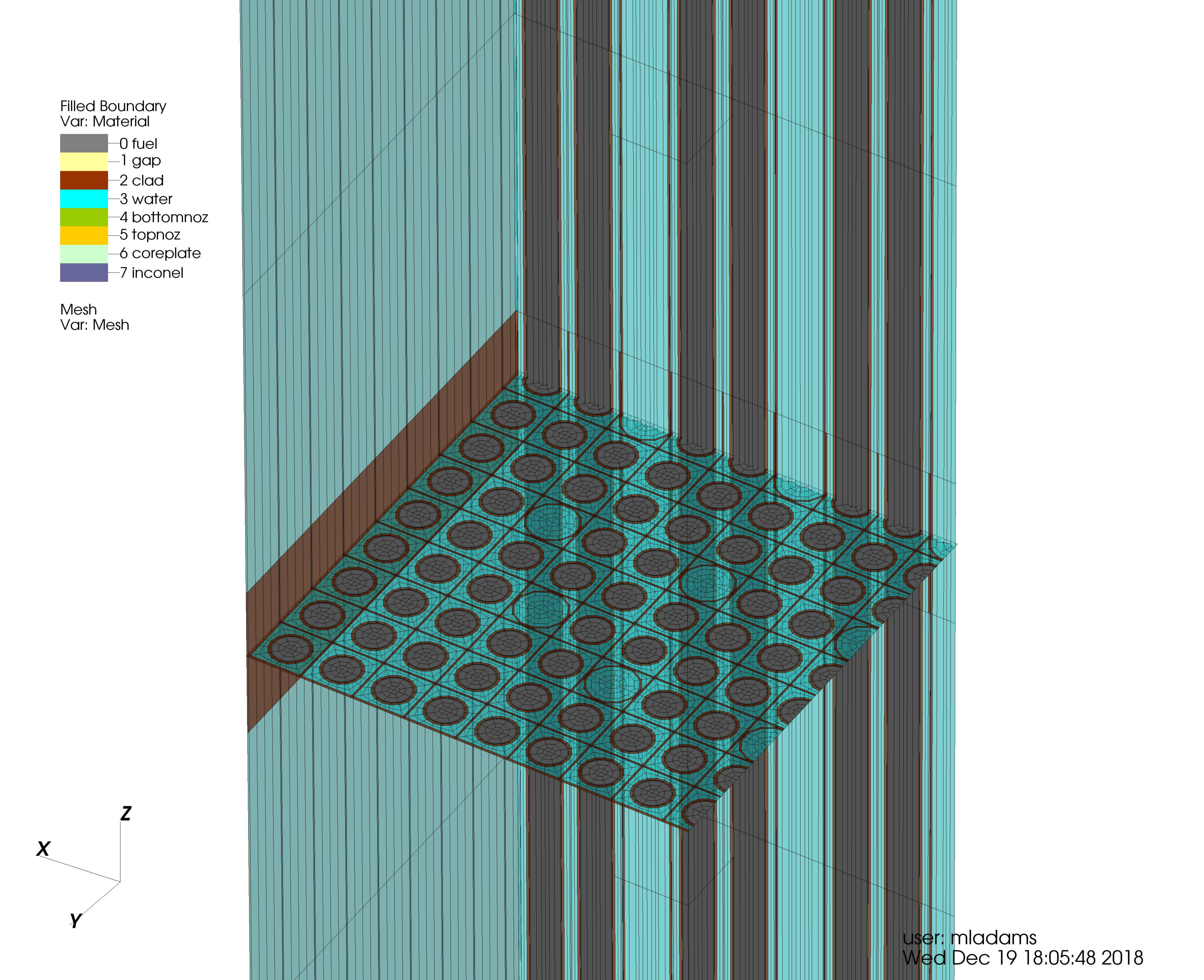}
\caption{Illustration of 3D mesh used in polygonal-prism sweep tests.  Left figure is not a CAD drawing, but is a cutout portion of PDT's polygonal-prism mesh with cells colored according to materials (zircaloy cladding and guide tubes, simplified bottom inconel grid spacer, and simplified bottom inlet nozzle).  Geometry and material properties, including simplifications, are as specified CASL VERA benchmark problem 3A \cite{verabenchmarks}.  Right figure is a 3-slice view of an axial segment of the mesh for a quarter of the problem 3A fuel assembly, with the xy slice going through a zircaloy grid spacer.  Figure~\ref{fig:2dzoom} shows details of the polygonal $xy$ mesh, which is extruded in $z$ to form polygonal prisms.  Visualizations are from the Visit code \cite{visit}.}
\label{fig:3dview}
\end{figure}

\begin{figure}[h!]
\centering
\includegraphics[width= 0.5\linewidth]{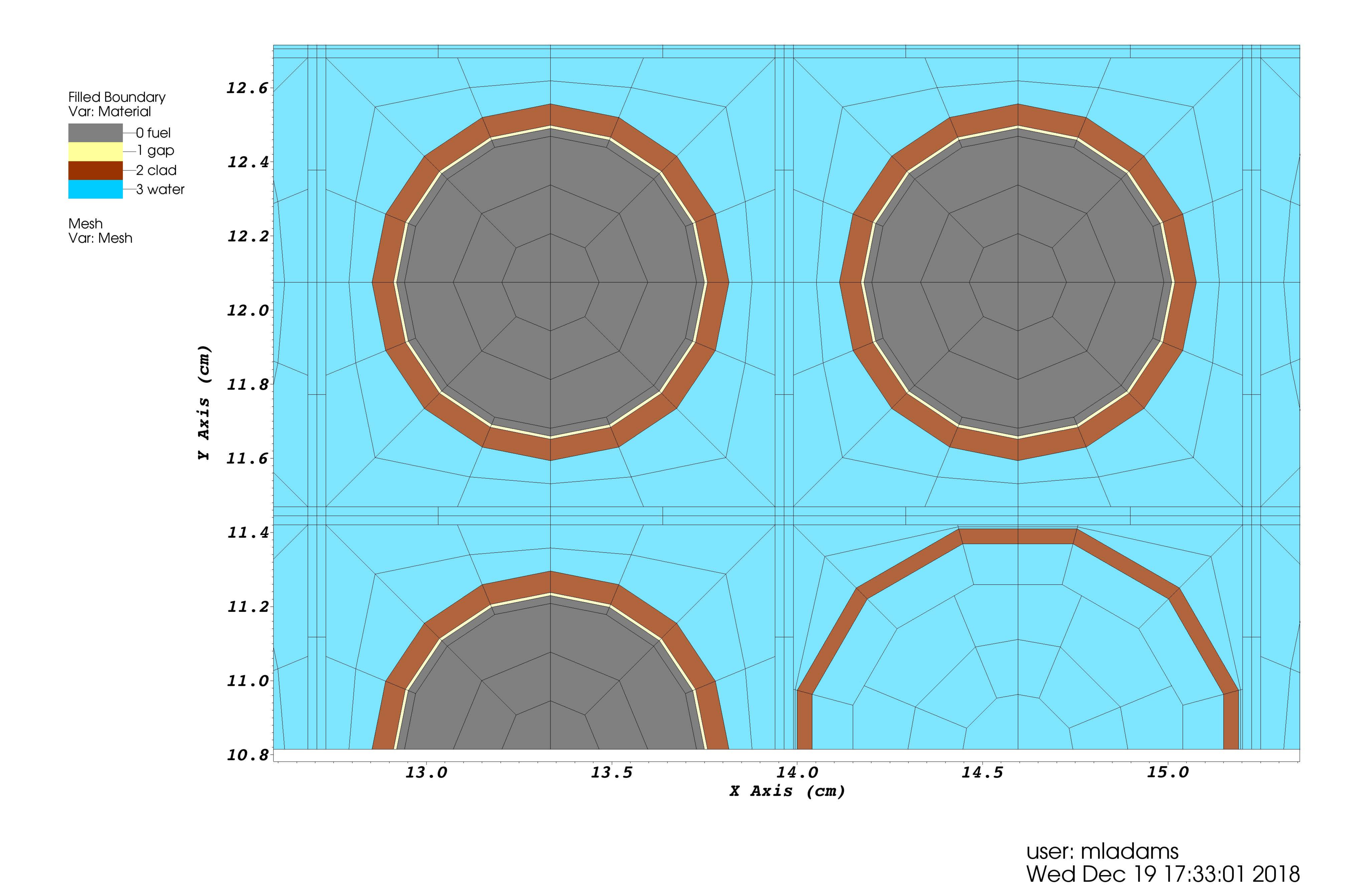}
\caption{Closeup of radial ($xy$) mesh in reactor test problem, showing quarters of 3 fuel pin cells and one guide tube.  The gap between fuel and cladding is resolved, as specified in the benchmark problem statement.  The rectangular cells on the pin-cell borders are filled with grid-spacer material in the specified axial intervals.}
\label{fig:2dzoom}
\end{figure}

Figures~\ref{fig:3dview} and~\ref{fig:2dzoom} illustrate the meshes used for testing our sweep methodology on polygonal-prism grids.  Our sweep tests used core counts ranging from 1,632 to 767,584.  As discussed previously, when we run one fourth of a problem using two reflecting boundaries, our 767,584-core results are essentially the results we would obtain if we ran the full problem with $4 \times 767,584 = 3,070,336$ cores.  To maintain consistency with previous results (which had no reflecting boundaries), we plot our two-reflecting-boundary performance results in this section as a function of ``effective'' core count, which is 4 times the actual core count.


In our study of sweeping on polygonal-prism grids we kept unknown count per core roughly the same as we scaled up in core count.  In all problems we used 65 energy groups, which we divided into three ``groupsets'' of 12, 31, and 22 groups, respectively.  This gave us three different sweep data points for each problem, because the sweeps were performed one groupset at a time.  In this study we added spatial cells by adding fuel assemblies, beginning with a reflected quarter-assembly and ramping up to a reflected 4$\times$4 array of assemblies (a factor of 64 in number of fuel rods), and also by increasing axial resolution by almost a factor of 2.  We also increased directional resolution by allowing quadrature sets to range from 64 directions/octant (low-energy groups, low resolution) to 768 directions/octant (high-energy groups, high resolution). Table~\ref{tab:rxscaling} provides details of the number of assemblies, axial cell count, and quadrature sets for each groupset, for each \textbf{full} problem in our test suite.  As discussed previously, we obtained our results using one-fourth of the indicated cores on one-fourth of the indicated full problems, with two reflecting boundaries.

\begin{table}[h!]
  \caption{\bf Test Problem Parameters, Polygonal-Prism Grids (full problem---see text)}
  \label{tab:rxscaling}
  \centering
  \def\arraystretch{1.0}
  \begin{tabular}{| r | r | r | r | r | r | r | r |} 
  \hline
     {\bf  } & {\bf  } & {\bf Axial}  & {\bf Total} & GS 1 & GS 2 & GS 3 & Total  \\ 
     {\bf Cores} & {\bf Assys} & {\bf cells}  & {\bf cells} & (12 grps) & (31 grps) & (22 grps) & unknowns  \\ 
     {\bf  } & {\bf  } & {\bf  }  & {\bf  } & directions & directions & directions & /core  \\ \hline
     6528 & 1$\times$1 & 96 & 3.3 E6 & 8$\times$6$\times$32  & 8$\times$6$\times$16 & 8$\times$8$\times$8 & 2.3 E8  \\  \hline
     27,744 & 2$\times$2 & 96 & 13.3 E6 & 8$\times$6$\times$32  & 8$\times$6$\times$16 & 8$\times$12$\times$8 & 2.4 E8  \\  \hline
     78,608 & 2$\times$2 & 136 & 1.9 E7 & 8$\times$12$\times$32  & 8$\times$12$\times$16 & 8$\times$12$\times$12 & 2.2 E8  \\  \hline
     314,432 & 4$\times$4 & 136 & 7.6 E7 & 8$\times$12$\times$32  & 8$\times$12$\times$16 & 8$\times$12$\times$16 & 2.4 E8  \\  \hline
     1,414,944 & 6$\times$6 & 136 & 1.7 E8 & 8$\times$16$\times$48  & 8$\times$12$\times$24 & 8$\times$24$\times$16 & 2.1 E8  \\  \hline
     3,070,336 & 8$\times$8 & 166 & 3.7 E8 & 8$\times$16$\times$48  & 8$\times$12$\times$24 & 8$\times$24$\times$16 & 2.1 E8  \\  \hline
\end{tabular}
\end{table}

\begin{figure}[h!]
\centering
\includegraphics[width= 0.75\linewidth]{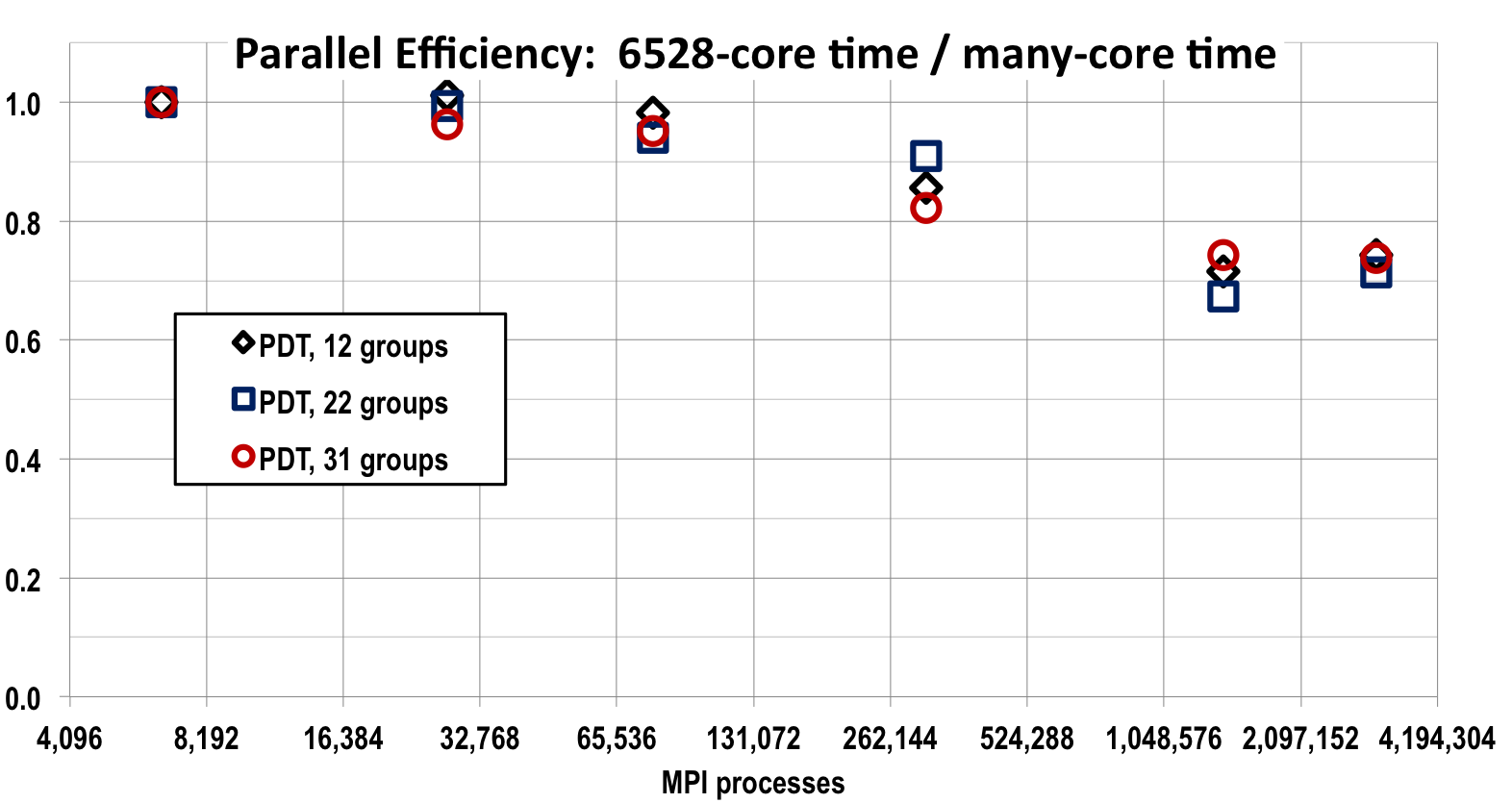}
\caption{Performance as a function of process count from optimal sweep algorithm on unstructured grids, using IBM BG/Q computers.  ``Efficiency'' is sweep-execution time per unknown per process on a given number of processes divided by the same on 6528 processes, with work per process approximately constant.  Different symbols show results for different ``groupsets.''  See text and Table~\ref{tab:rxscaling} for details.}
\label{fig:rxscaling}
\end{figure}

Results are shown in Fig.~\ref{fig:rxscaling}, normalized to the single-assembly problem, which used 6528 cores with one MPI process per core.  Results are in terms of ``grind times," which are defined to be time per sweep per unknown per core.  Each data point is a grind time at 6528 cores divided by grind time at the indicated core count.  Three different sets of points are plotted in the Figure---one for each groupset.  In the PDT code, a task is a set of cells (cellset), a set of directions (angleset), and a set of groups (groupset).  The work function that executes a task must prepare for the task (reading angular fluxes from upstream cellsets) and loop through the cells in the appropriate order for the given set of directions.  For each cell there is a loop over directions in the angleset, and for each direction there is an innermost loop over groups in the groupset.  Inside the inner loop an $N \times N$ linear system is solved for the PWLD angular fluxes, where $N$ is the number of spatial degrees of freedom in the particular cell being solved.  With PWLD, $N$ is the number of vertices in the polyhedral cell.  Because of the nesting of the loops, larger anglesets and groupsets produce lower grind times, if all else is equal, because the work done preparing for the task and pulling in cellwise information is amortized over the calculation of more unknowns.  This is why the results differ for the different groupsets---they have different numbers of groups and different numbers of directions per angleset.

%% file: s6-conclusions.tex
\section{CONCLUSIONS}

Sweeps can be executed efficiently at high core counts.  One key to achieving 
efficient performance is an optimal
scheduling algorithm that executes simultaneous multi-octant sweeps with the
minimum possible idle time.  Another is partitioning and aggregation factors
that minimize total sweep time.  An ingredient that helps to attain this is a
performance model that predicts performance with reasonable quantitative 
accuracy.  Of course, none of this is sufficient to attain excellent parallel 
efficiency without great care in implementation.  But with all of these 
ingredients in place, sweeps can be executed with high efficiency beyond $10^6$ 
concurrent processes.

Our computational results demonstrate this.  They also show that at
least two different sweep scheduling algorithms achieve the minimum possible
stage count, in agreement with our theory and ``proof."  The common
perception that sweeps do not scale beyond a few thousand cores is simply 
not correct.  Even with a relatively small problem (3 energy groups, 80 total
directions, and 4096 cells per core) our PDT/STAPL code has achieved
approximately 67\% efficiency with 1.57 $\times 10^6$ MPI processes, 
relative to an 8-process calculation, and the ARDRA code has achieved 71\% efficiency (total solve time) at the same process count on a problem with more energy groups and directions.  With additional energy groups and
directions, parallel efficiency improves further.  We have reason to believe
that further refinement of some implementation details will increase the efficiencies reported here.

The analysis and results in this summary are for 3D Cartesian grids with ``brick"  
cells and for certain grids that are unstructured at a fine scale but structured at a 
coarse scale.  To illustrate the latter kind of grid we have shown results here from 
a series of nuclear-reactor calculations whose grids resolve complicated geometries with 
high fidelity.  We are also working on sweeps for AMR-type grids, arbitrary 
polyhedral-cell grids without a coarse structure, and grids for which it is difficult to achieve load balancing.  We plan to present results in a future communication.

In this paper we have restricted our attention to spatial domain decomposition 
with $P_x \times P_y \times P_z$ partitioning, in which each processor owns a 
brick-shaped contiguous subdomain of the spatial domain.  For some grids and 
problems there may be efficiency gains if processors are allowed to ``own" 
non-contiguous collections of cellsets, an option considered in \cite{mc2015-sweep} and \cite{BaileyFalgout}, with the terminology ``domain overloading."  We expect to report on this family of partitioning and aggregation methods in the future.

In this paper we have restricted our attention to 

Reflecting boundaries introduce direction-to-direction dependencies that 
decrease available parallelism.  We have shown that with our sweep algorithm, 
the parallel solution with $P$ processors on a problem with $n$ mutually orthogonal 
reflecting boundaries performs with the same efficiency as the parallel solution with 
$2^n \times P$ processors on the full domain without reflecting boundaries.

Curvilinear coordinates introduce a different kind of direction-to-direction dependency, again reducing available parallelism and probably making sweeps somewhat less efficient than in Cartesian coordinates.  We have not yet devoted much attention to parallel sweeps in curvilinear coordinates, but we expect to address this in the future.

%% file: s7-references.tex
\setlength{\baselineskip}{12pt}

%% file: s8-appendix1.tex
\section{APPENDIX:  EXAMPLE $P_z=1$ or 2 SWEEP}
\label{sec:appendix1}

While the stage algebra in Sec.~\ref{sec:proofs} is necessary for our proofs, a
visual illustration makes the actual behavior of our algorithms much more
accessible.  We begin with an extremely simple example, a 2D sweep with $M=3$.
Note:  This behavior is identical to that for $P_z=1$ or for either half of the
processes for $P_z=2$.

\begin{figure}[h!]
\centering
\includegraphics[width=1.0\linewidth]{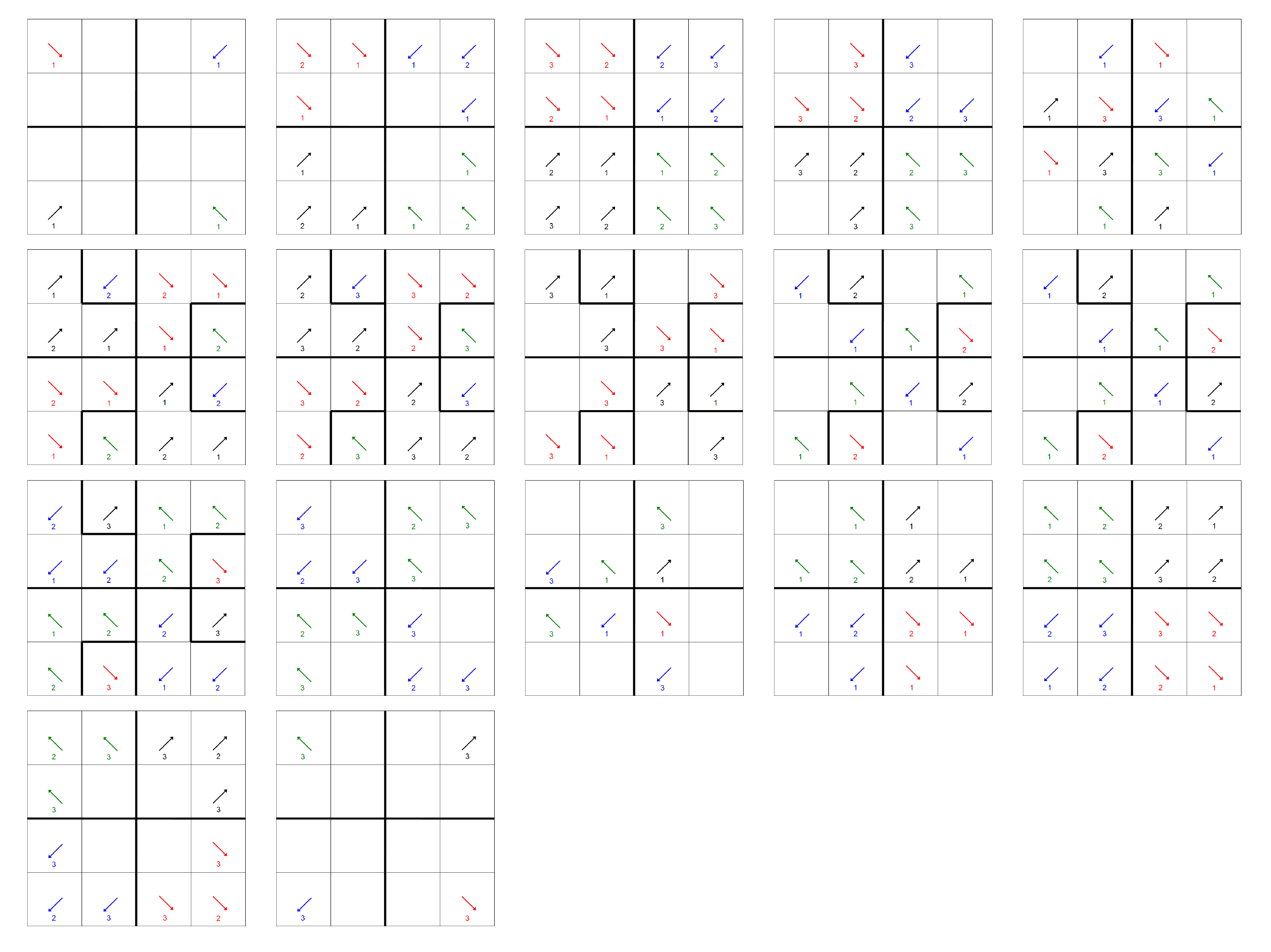}
\caption{{\bf Example sweep.}  Stages progress left to right and top to bottom.  Anglesets in the same quadrant have arrows of the same color.  Example shows three anglesets per quadrant.}
\label{fig:2d-sweep}
\end{figure}

For our 2D example, we have illustrated the behavior of the ``depth-of-graph''
algorithm.  In the next appendix, we present a 3D sweep using the
``push-to-central'' algorithm.  In Fig.~\ref{fig:2d-sweep}, successive stages
are presented first from left to right and then from top to bottom.  The
16-process layout presented could be either the full domain or only the center
of a larger domain; the behavior is the same whether there are processes outside
of this range or not.  Bold lines depict boundaries between priority regions.

%% file: s9-appendix2.tex
\section{APPENDIX:  EXAMPLE 3D VOLUMETRIC SWEEP}
\label{sec:appendix2}

\begin{figure}[!htb]
\centering
\includegraphics[width=1.385\linewidth,angle=90]{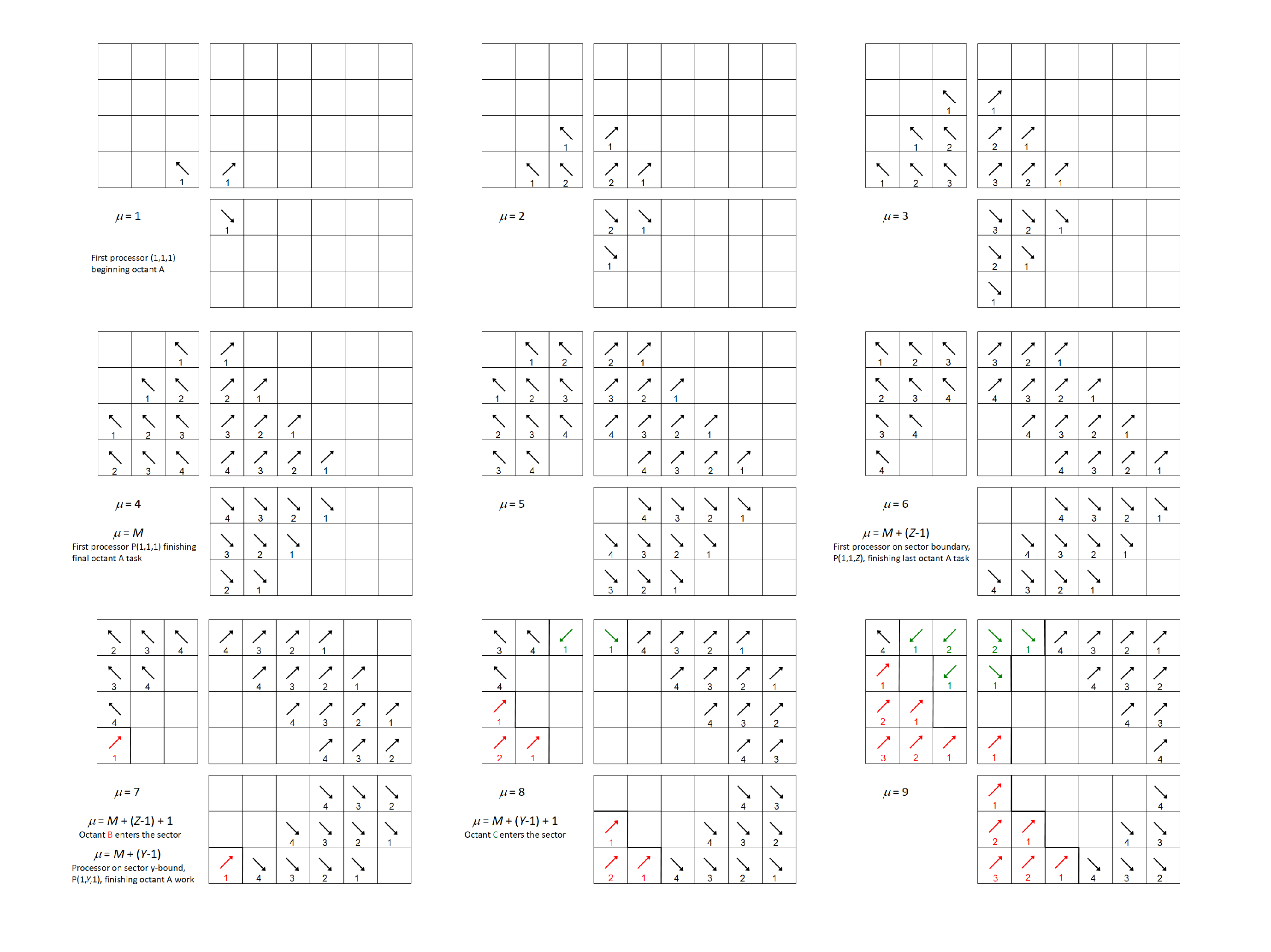}
\vspace{-12mm}
\caption{{\bf Stages 1-9.}}
\label{fig:stages-1-9}
\end{figure}

\begin{figure}[!htb]
\centering
\includegraphics[width=1.385\linewidth,angle=90]{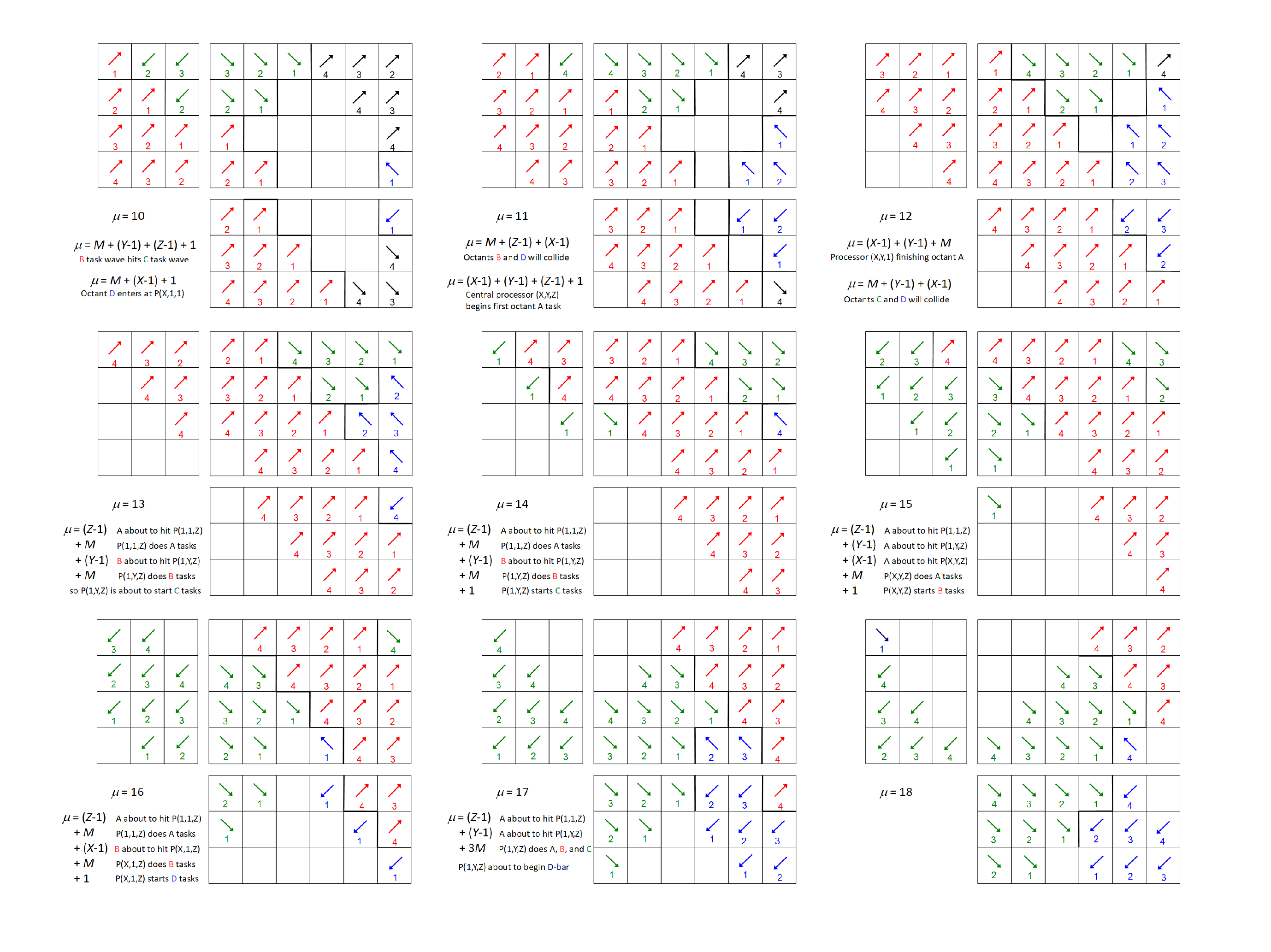}
\vspace{-12mm}
\caption{{\bf Stages 10-18.}}
\label{fig:stages-10-18}
\end{figure}

\begin{figure}[!htb]
\centering
\includegraphics[width=1.385\linewidth,angle=90]{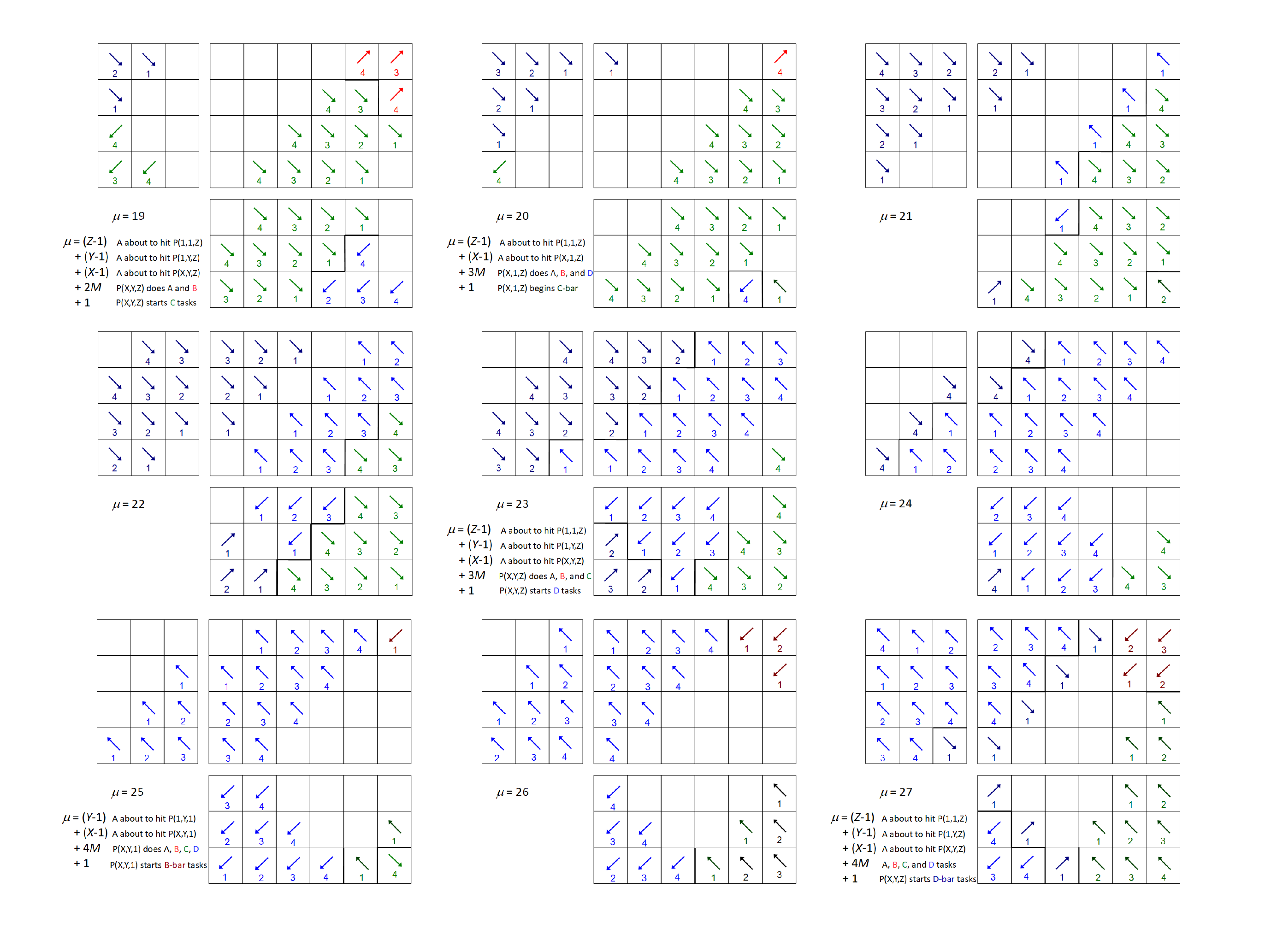}
\vspace{-12mm}
\caption{{\bf Stages 19-27.}}
\label{fig:stages-19-27}
\end{figure}

\begin{figure}[!htb]
\centering
\includegraphics[width=1.385\linewidth,angle=90]{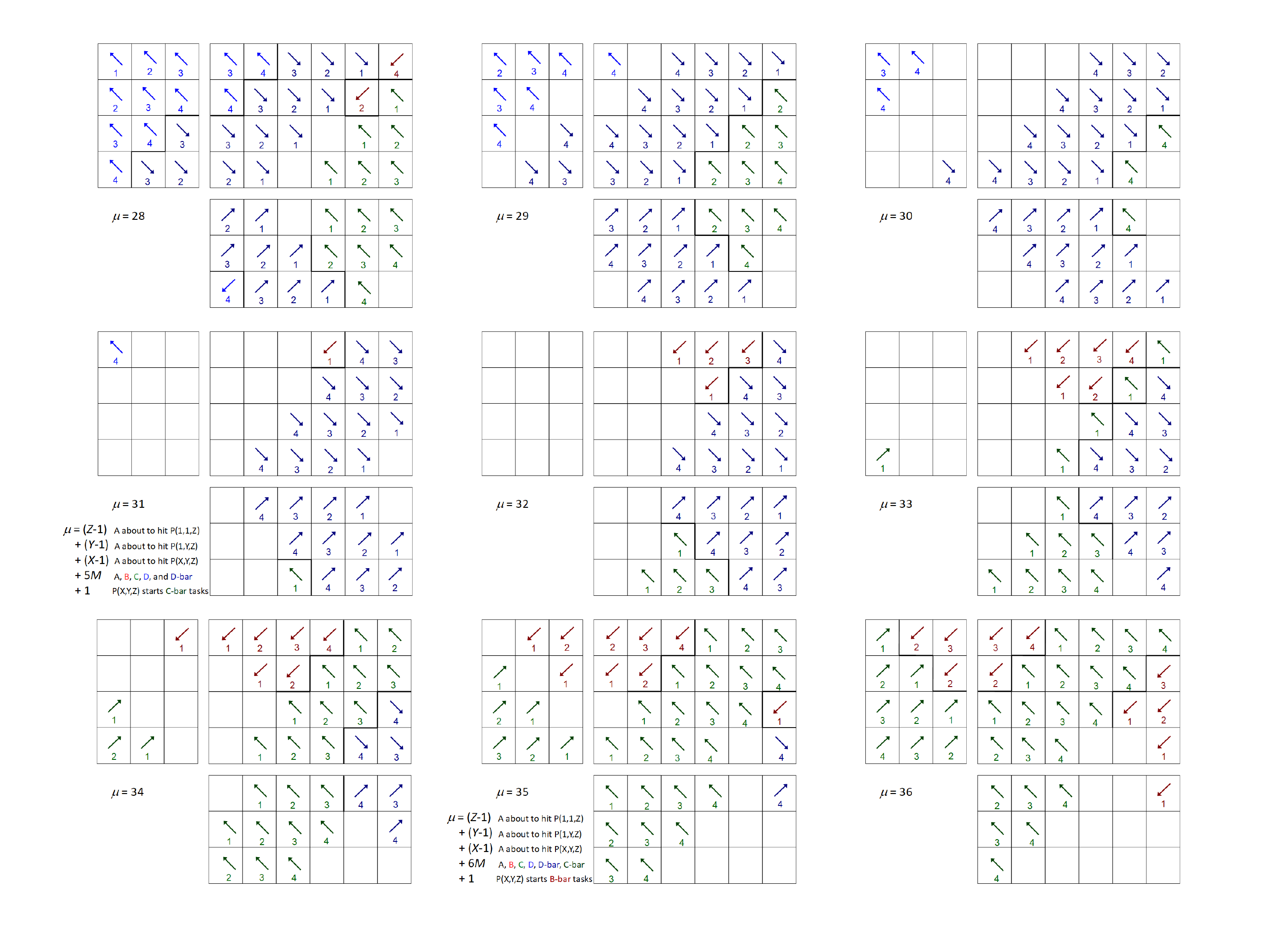}
\vspace{-12mm}
\caption{{\bf Stages 28-36.}}
\label{fig:stages-28-36}
\end{figure}

\begin{figure}[!htb]
\centering
\includegraphics[width=1.385\linewidth,angle=90]{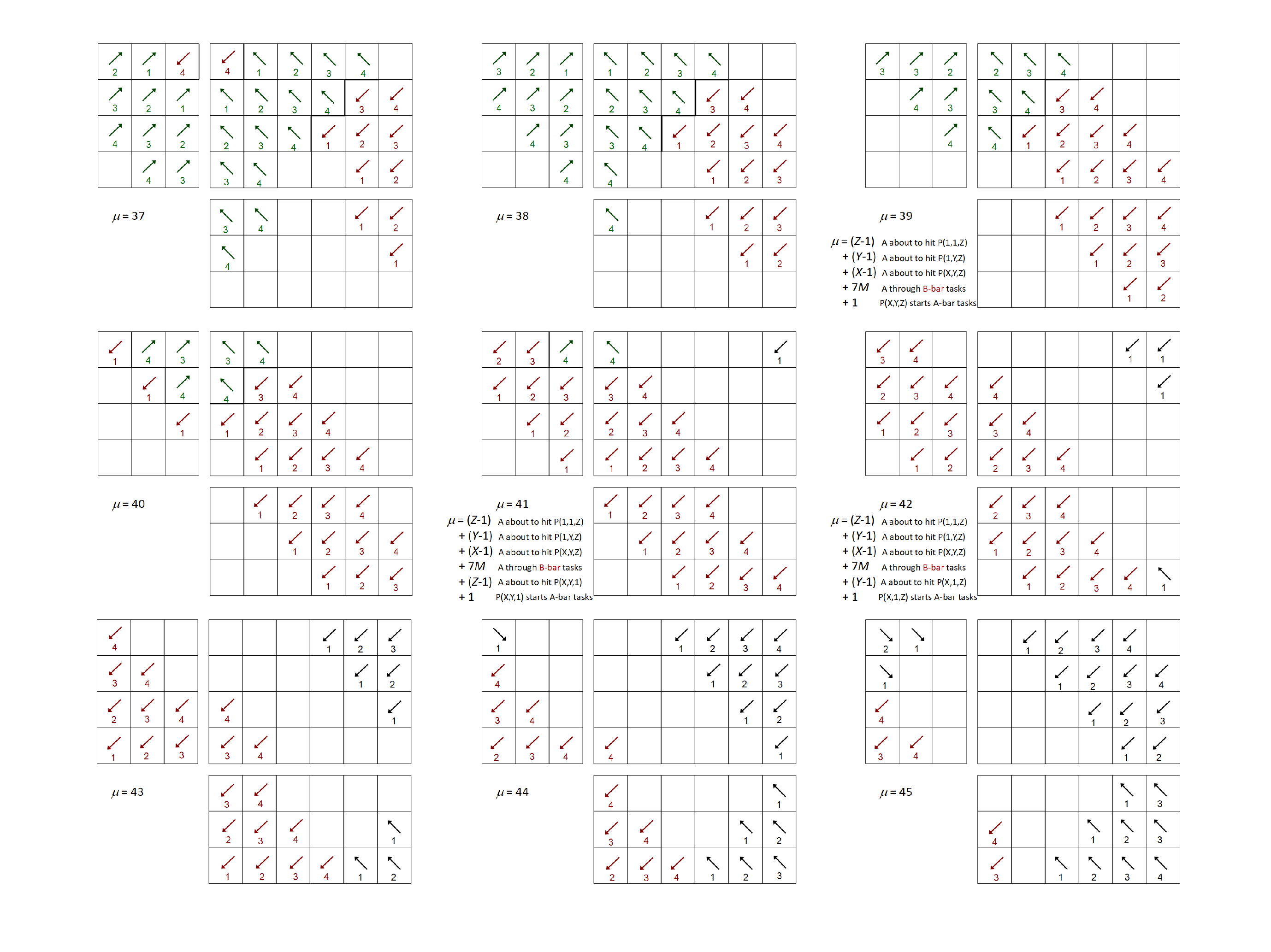}
\vspace{-12mm}
\caption{{\bf Stages 37-45.}}
\label{fig:stages-37-45}
\end{figure}

\begin{figure}[!htb]
\centering
\includegraphics[width=1.385\linewidth,angle=90]{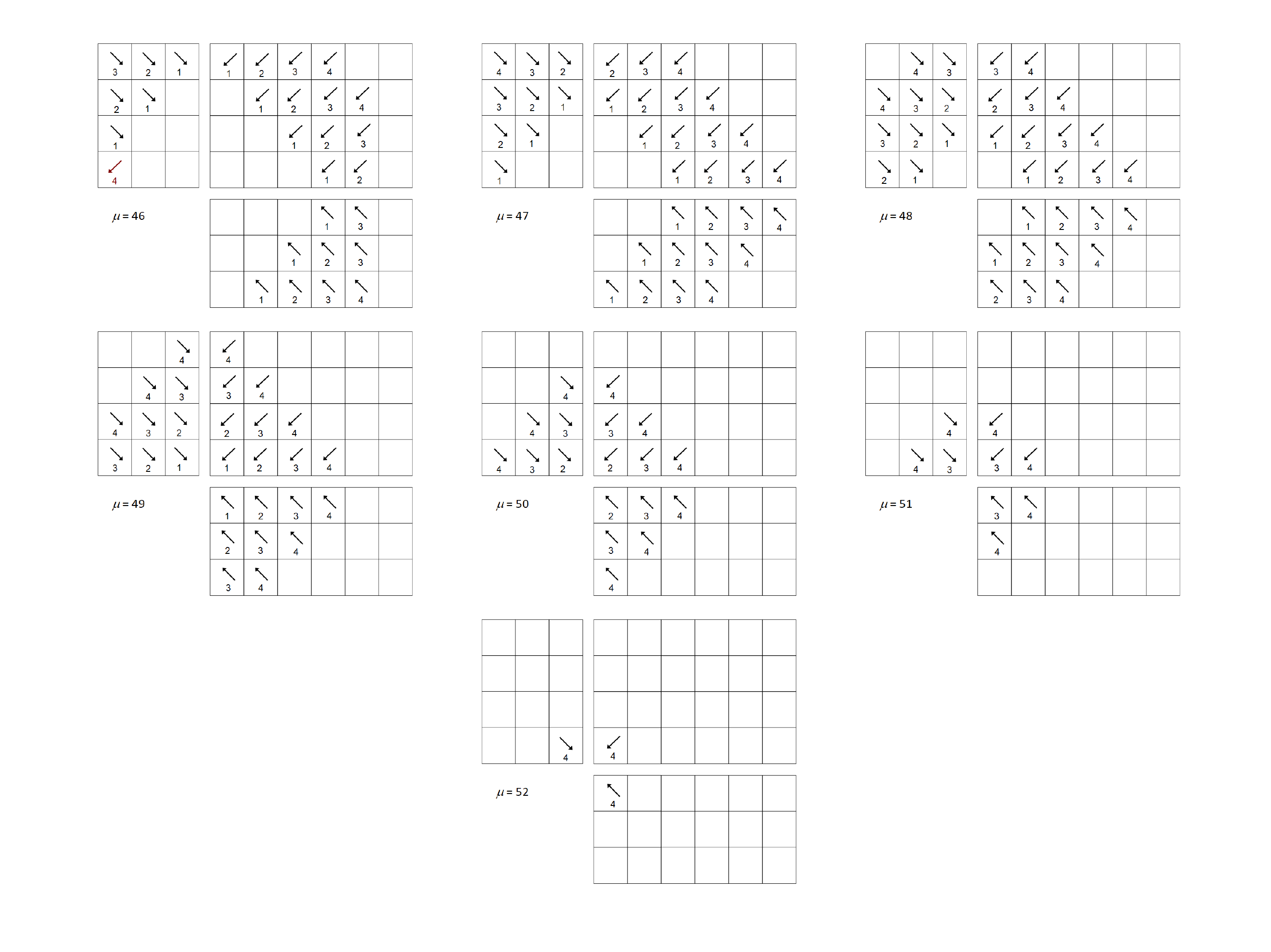}
\vspace{-12mm}
\caption{{\bf Stages 46-52.}}
\label{fig:stages-46-52}
\end{figure}

Figures~\ref{fig:stages-1-9}-\ref{fig:stages-46-52} illustrate an example sweep.
In the example, there are four anglesets per octant and 576 processes, with
$P_x=12$, $P_y=8$, and $P_z=6$.  We illustrate the ``push-to-central''
algorithm; i.e., this behavior is different from that described in
Sec.~\ref{sec:proofs} in terms of priority regions.  Here, the entire sector
shares the same priority ordering, specifically $(A, B, C, D, \bar D, \bar C,
\bar B, \bar A)$.

\begin{figure}[!htb]
\centering
\vspace{-4mm}
\includegraphics[scale=0.5]{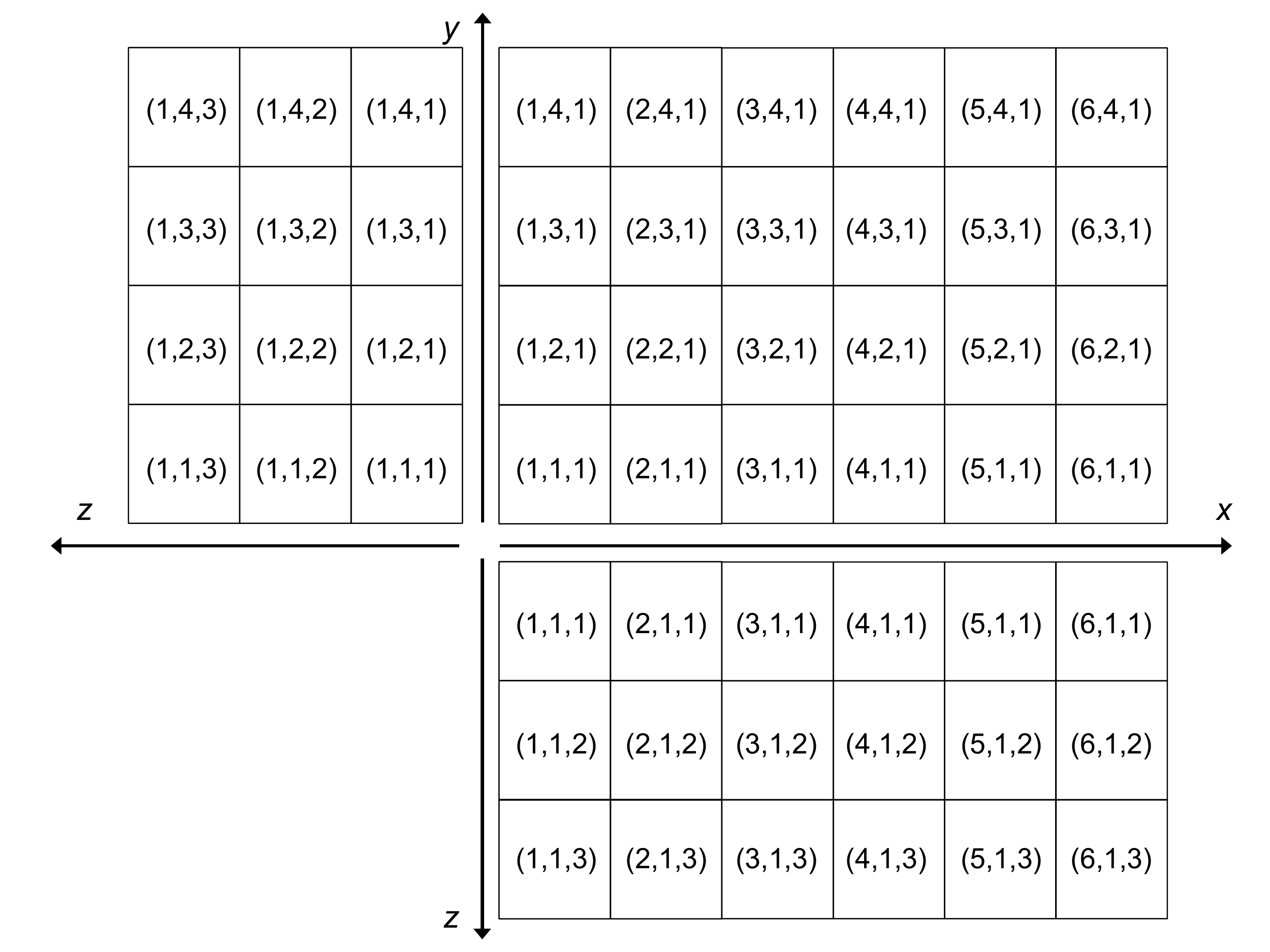}
\vspace{-5mm}
\caption{{\bf Open Box Diagrams.}  Each panel represents a planar ``slice'' of
processes.  The two axes adjacent to each panel define its orientation.
$(i,j,k)$ indices are shown for each process.}
\label{fig:open-box}
\end{figure}

We show the order of task execution for processes with $P_x \in (1,X)$, $P_y
\in (1,Y)$, and $P_z \in (1,Z)$ using what we call ``open box'' diagrams (see
Fig.~\ref{fig:open-box}).  The diagrams show the sets of processes in the
region with $P_x=1$ (top right), $P_y=1$ (bottom right), and $P_z=1$ (top left).
Tasks within a given octant are numbered from $1-4$ and are shown with arrows
representing the directions of dependencies.  (The arrows may appear to have
different directions on different panels; this is because each panel has its own
orientation.)  They're also color coded for clarity.

This being the ``push-to-central'' algorithm, the collisions between task waves
are not static as they are for the ``depth-of-graph'' algorithm.  Rather, task
waves of higher priority overtake the lower priority waves, which lie dormant
until they are able to re-emerge from the trailing end of the priority wave.
This happens between nearly every pair of octants.  Here, then, the bold lines
represent sweepfront collisions, not priority region boundaries as in the 2D
example sweep.

Stage counts are included in the figure, as well as occasional notes pointing
out salient features in the behavior of the sweep algorithm and their connection
with stage counts.  Again, the requirements for optimality are simply that the
central processes begin working at the first possible stage, that they stay
busy until their tasks are finished, and that the final octant's tasks proceed
uninterrupted to the boundary.  To that end, the figures note the stages when
process $P(X,Y,Z)$ begins each octant.  In this example, the optimal stage
count is $P_x + P_y + P_z + 8M = 52$, which is indeed achieved by the
``push-to-central'' algorithm seen below.

